\documentclass[12pt]{article}
\pdfoutput=1

\usepackage[utf8]{inputenc}
\usepackage[left=2.55cm, right=2.55cm, top=2.55cm, bottom=2.55cm]{geometry}
\usepackage{amsmath,amssymb,amsbsy}
\usepackage{slashed}
\usepackage{xcolor}
\usepackage{graphicx}
\usepackage{cancel}
\usepackage{url}
\usepackage{cancel}
\usepackage{comment}
\usepackage{tcolorbox} 
\usepackage{cite}
\usepackage[colorlinks=true,allcolors=darkpurple,pdfborder={0 0 0},linktocpage=false]{hyperref}
\usepackage{tabularx,booktabs}
\usepackage{multicol}
\usepackage{units}
\usepackage{xspace}
\usepackage[labelfont=bf]{caption}
\usepackage[section]{placeins}
\usepackage{enumitem}
\usepackage{ulem}
\usepackage{subcaption}
\usepackage{multirow}
\usepackage{comment}
\usepackage{verbatim}
\usepackage{ulem}
\useunder{\uline}{\ul}{}
\usepackage{tensor}
\usepackage{slashed}
\usepackage{array} 
\newcolumntype{C}[1]{>{\centering\arraybackslash}p{#1}}
\renewcommand{\arraystretch}{1.5}
\usepackage[compat=1.1.0]{tikz-feynman}
\tikzfeynmanset{warn luatex=false}

\definecolor{darkred}{rgb}{0.6,0,0}
\definecolor{darkpurple}{rgb}{0.5,0,0.5}
\definecolor{lightblue}{rgb}{0.93, 0.95, 1.0}
\definecolor{mygray}{rgb}{0.9,0.9,0.9}
\definecolor{avblue}{rgb}{0.0, 0.0, 0.8}
\definecolor{javi}{rgb}{0.5,0,1}
\definecolor{asparagus}{rgb}{0.53, 0.66, 0.42}
\definecolor{aqua}{rgb}{0.4, 0.6, 0.7}

\def\abs[#1]{\lvert #1\rvert}
\def\L{\mathcal{L}}
\def\M{\mathcal{M}}
\def\S{\mathcal{S}}
\def\P{\mathcal{P}}

\def\hc{\text{h.c.}}
\def\BR{\text{BR}}

\def\z2{$\mathbb{Z}_2$}
\def\Tr{\text{Tr}}

\def\id{\mathbb{I}}

\def\U1L{$\mathrm{U(1)}_L$} 
\def\3221{$SU(3)_c \times SU(2)_1 \times SU(2)_2 \times U(1)_Y$}
\def\221{$SU(2)_1 \times SU(2)_2 \times U(1)_Y$}
\def\22{$SU(2)_1 \times SU(2)_2$}
\def\55{$SU(5)_1 \times SU(5)_2$}

\newcommand{\javi}[1]{{\color{javi} \bf #1}}

\usetikzlibrary{shapes.geometric, fit}

\definecolor{seeI}{RGB}{35,110,215}    
\definecolor{seeL}{RGB}{30,150,70}     
\definecolor{seeInv}{RGB}{210,40,40}   

\newcommand{\seesaw}[2]{%
  \begin{tikzpicture}[baseline=(t.base)]
    \node[ellipse, draw=#1, line width=0.8pt,
          inner xsep=0pt, inner ysep=1pt] (t) {$\displaystyle #2$};
  \end{tikzpicture}}

\newcommand{\seesawdouble}[3]{%
  \begin{tikzpicture}[baseline=(t.base)]
    \node[ellipse, draw=#1, line width=0.8pt,
          inner xsep=0pt, inner ysep=1pt] (t) {$\displaystyle #3$};
    \node[ellipse, draw=#2, line width=0.8pt, inner sep=1.8pt, fit=(t)] {};
  \end{tikzpicture}}

\newcommand{\typeI}[1]{\seesaw{seeI}{#1}}
\newcommand{\linear}[1]{\seesaw{seeL}{#1}}
\newcommand{\inverse}[1]{\seesaw{seeInv}{#1}}
\newcommand{\linInv}[1]{\seesawdouble{seeInv}{seeL}{#1}}

\newcommand{\cunderbrace}[3]{%
  {\color{#1}\underbrace{\normalcolor #2}_{\text{#3}}}}

\newcommand {\ignore}[1]{}

\definecolor{pendingcol}{rgb}{0.80,0.33,0.0}
\newcommand{\pending}[1]{{\color{pendingcol}\sffamily\bfseries [PENDIENTE: #1]}}

\newcommand{\AddrIFIC}{%
  Instituto de F\'{i}sica Corpuscular, CSIC-Universitat de Val\`{e}ncia, 46980 Paterna, Spain}

\newcommand{\AddrFISTEO}{%
  Departament de F\'{\i}sica Te\`{o}rica, Universitat de Val\`{e}ncia, 46100 Burjassot, Spain}

\begin{document}

\begin{center}
\vspace*{15mm}

\vspace{1cm}
{\Large \bf 
Hybrid seesaw from $\boldsymbol{SU(2)_1 \times SU(2)_2 \times U(1)_Y}$
} \\
\vspace{1cm}

{\bf Antonio Herrero-Brocal$^{\text{a}}$, Javier Perez-Soler$^{\text{a}}$, Avelino Vicente$^{\text{a,b}}$}

\vspace*{.5cm}
 $^{(\text{a})}$ \AddrIFIC \\\vspace*{.2cm} 
 $^{(\text{b})}$ \AddrFISTEO

\vspace*{.3cm}
\href{mailto:antonio.herrero@ific.uv.es}{antonio.herrero@ific.uv.es},
\href{mailto:javier.perez.soler@ific.uv.es}{javier.perez.soler@ific.uv.es},
\href{mailto:avelino.vicente@ific.uv.es}{avelino.vicente@ific.uv.es}
\end{center}

\vspace*{10mm}
\begin{abstract}\noindent\normalsize
We study neutrino mass generation in a model based on the $SU(3)_c \times SU(2)_1 \times SU(2)_2 \times U(1)_Y$ gauge symmetry, in which the Standard Model $SU(2)_L$ arises at low energies as the diagonal subgroup of the two
$SU(2)$ factors. The symmetry is enlarged with a global $U(1)_{B-L}$ symmetry, spontaneously broken by the vacuum expectation value of a scalar bidoublet. Light neutrino masses are generated at tree level through a hybrid of the type-III, inverse and linear seesaw mechanisms. A distinctive feature of the construction is the existence of a Nambu-Goldstone boson, the majoron, which transforms non-trivially under the original gauge symmetry while being mostly singlet under the Standard Model gauge group. We explore the phenomenological consequences of this setup, including electroweak precision constraints, the production of the nearly degenerate $W'$ and $Z'$ bosons at colliders and lepton flavor violating decays with a majoron in the final state. The latter constitute the most promising signature of the model.
\end{abstract}

\section{Introduction}
\label{sec:Intro}

The observation of neutrino oscillations established that neutrinos are massive, providing the first laboratory evidence of physics beyond the Standard Model (SM)~\cite{Kajita:2016cak,McDonald:2016ixn}. In the SM, neutrinos are strictly massless: the model contains no right-handed
neutrino singlets and $B-L$ is an accidental symmetry of the renormalizable
Lagrangian, so that no Majorana mass term can be written either. Accommodating the oscillation data therefore requires extending the SM with new
fields and, possibly, with new interactions as well. The way in which this is
done is far from unique, and the resulting scenarios differ not only in the
scale at which the new physics lies, but also in whether $B-L$ is
violated explicitly or spontaneously, and in the observables that may reveal it.

The gauge symmetry of a model has a strong impact on the way neutrino masses are
generated. Extending the SM gauge group not only constrains the operators allowed in the Lagrangian, but also enlarges the field content. The classic examples are left-right symmetric
models~\cite{Pati:1974yy,Mohapatra:1974hk,Senjanovic:1975rk,Mohapatra:1979ia},
in which the right-handed neutrino is not an optional addition but a necessary
component of a lepton doublet under $SU(2)_R$, and $B-L$
extensions~\cite{Davidson:1978pm,Marshak:1979fm,Mohapatra:1980qe}, where anomaly
cancellation requires three SM singlet fermions that naturally play the role of
the heavy states of the type-I seesaw. Beyond these standard scenarios, more exotic gauge extensions have also been explored, leading to a variety of mass generation mechanisms~\cite{Singer:1980sw,Babu:1985gi,He:1990pn,Pisano:1992bxx,Frampton:1992wt,Boucenna:2015zwa,Herrero-Brocal:2025rkx}.

In this work we consider a model based on the \3221 gauge symmetry, which we will
denote as 3221 for short, with
\begin{align}
SU(2)_1 \times SU(2)_2 \,\, \to \,\, SU(2)_{\rm diag} \equiv SU(2)_L \, .
\end{align}
Neither of the two $SU(2)$ factors of the original gauge group can therefore be
identified with $SU(2)_L$, which instead emerges at low energies as their diagonal subgroup. This corresponds to the so-called breaking pattern II in the general classification of 3221 models of Ref.~\cite{Hsieh:2010zr}. This gauge structure is reminiscent of a deconstructed electroweak sector, where the diagonal subgroup surviving at low energies plays the role of the SM gauge group~\cite{ArkaniHamed:2001nc,ArkaniHamed:2001ky}. Compared to its left-right cousin, this class of models has received only limited attention as far as the generation of neutrino masses is concerned~\cite{Hue:2016nya,CarcamoHernandez:2025eyt,Leite:2025yuq}. Our goal is to study a simple 3221 model with spontaneous $B-L$ violation that realizes a hybrid between the type-III~\cite{Foot:1988aq}, inverse~\cite{Mohapatra:1986bd} and linear seesaw~\cite{Akhmedov:1995ip,Akhmedov:1995vm} mechanisms. A distinctive feature of our construction is the appearance of a massless Nambu-Goldstone boson (NGB), the majoron~\cite{Chikashige:1980qk,Chikashige:1980ui,Schechter:1981cv,Gelmini:1980re,Aulakh:1982yn}, which transforms non-trivially under the original 3221 symmetry while being mostly singlet under the SM gauge group. As we will show, this state opens up a rich set of experimental signatures that can be used to probe the scenario.

The rest of the paper is organized as follows. In Sec.~\ref{sec:Model}, we introduce the model and discuss its main properties. In Sec.~\ref{sec:Spectrum}, we analyze the particle spectrum and derive exact or approximate expressions for the masses of all states. Sec.~\ref{sec:Pheno} is devoted to the phenomenology, with particular emphasis on the features specific to our setup. We summarize our results and present our conclusions in Sec.~\ref{sec:Summ}, while technical details are collected in Appendices~\ref{App:Symm}, \ref{App:Seesaw}, \ref{App:Scalar} and \ref{App:SJJ}.
\section{The model} \label{sec:Model}
In this work we consider an extension of the electroweak group to $SU(2)_1 \times SU(2)_2\times U(1)_Y$. In addition, we introduce a global $U(1)_{B-L}^{\text{global}} \equiv U(1)_{B-L}$ symmetry, leading to the full electroweak symmetry group
\begin{equation}
SU(2)_1 \times SU(2)_2\times U(1)_Y \times U(1)_{B-L} \, .
\end{equation}

To begin with, let us clarify why we choose $B-L$, instead of $L$ as the global symmetry. Although imposing $U(1)_L$ would lead to the same Lagrangian, $B-L$ provides a more transparent interpretation, as it is an exact symmetry of the theory, whereas $L$ is broken by anomalies. As shown in Ref.~\cite{Herrero-Brocal:2026nmc}, when a single VEV breaks several global $U(1)$ combinations, the associated NGB corresponds to the anomaly-free combination. Choosing $B-L$ therefore makes explicit that our majoron is an exactly massless NGB and that the model is free of the domain-wall problem~\cite{Herrero-Brocal:2026nmc}.

The gauge couplings and bosons of extended gauge sector are given by
\begin{align}
 SU(2)_1 : g_1 \, , \,  W_1^a &&   SU(2)_2 : g_2 \, , \,  W_2^a && U(1)_Y : g_Y \, , \,  B
\end{align}
where, as usual, $g_i$, $W_i^a$ and $B$ denote the coupling constants and gauge bosons, respectively. Here, $a$ represents the $SU(2)$ index, and Lorentz indices are omitted for simplicity.

We assume that all SM fields transform with the usual charges under $SU(2)_1\times U(1)_Y$, while they are singlets of $SU(2)_2$. Therefore, only the beyond the SM (BSM) fields are charged under the second $SU(2)$ group, $SU(2)_2$. In the scalar sector, we introduce an $SU(2)_2$ doublet $\eta$ and a complex bidoublet $\Phi$, carrying charges $1$ and $-1$ under $U(1)_{B-L}$, respectively. In component form, these fields are written as
\begin{equation}
    H=\left(\begin{array}{c}
         H_+  \\
         H_0 
         \end{array}\right)\;,\qquad
         \eta=\left(\begin{array}{c}
         \eta_+  \\
         \eta_0 
         \end{array}\right)\;,\qquad\Phi=\left(\begin{array}{cc}
        w_0 & x_+ \\
        y_- & z_0
    \end{array}\right)\;,
\end{equation}
where, in $\Phi$, the row and column indices correspond to $SU(2)_1$ and $SU(2)_2$, respectively.

In the fermionic sector, we introduce the field $N$, singlet under $SU(2)_1 \times SU(2)_2 \times U(1)_Y$ carrying charge $-1$ under $U(1)_{B-L}$, and a real bidoublet (therefore neutral under $U(1)_{B-L}$), denoted by $\chi$,~\footnote{In our treatment of a fermion bidoublet we follow the conventions of~\cite{Garcia-Cely:2015quu}.} and defined by
\begin{equation}
\chi \equiv \chi_R= \begin{pmatrix}
\chi_0 & \chi^+\\
\chi^- &  -\chi_0^c
\end{pmatrix}_R \, ,
\end{equation}
with $\chi^- = \left(\chi^+\right)^*$.  We consider $n_N$ and $n_\chi$ generations of $N$ and $\chi$, respectively. These numbers will be fixed below, once the neutrino mass matrix is computed, aiming for the most economical choice that accommodates oscillation data, in the spirit of~\cite{Abada:2014vea}. We summarize the relevant charges of the fields in the model in Tab.~\ref{table:particleContent}.
\setlength{\tabcolsep}{1em} 
\renewcommand{\arraystretch}{1.5}
\begin{table}[t!]
\centering
\begin{tabular}[t!]{||c c c c c c||} 
 \hline
 Field & Generations & $SU(2)_1$ & $SU(2)_2$ & $U(1)_Y$ & $U(1)_{B-L}$ \\ [0.5ex] 
 \hline\hline
 $L$ & 3 & $\mathbf{2}$ & $\mathbf{1}$ & $-1/2$ & $-1$ \\ 
 $e_R$ & 3 & $\mathbf{1}$ & $\mathbf{1}$ & $-1$ & $-1$ \\
 $N$ & $n_N$ & $\mathbf{1}$ & $\mathbf{1}$ & $0$ & $-1$ \\
 $\chi$ & $n_\chi$ & $\mathbf{2}$ & $\overline{\mathbf{2}}$ & $0$ & $0$ \\ [1ex] 
 \hline\hline
 $H$ & 1 & $\mathbf{2}$ & $\mathbf{1}$ & $1/2$ & $0$ \\
 $\eta$ & 1 & $\mathbf{1}$ & $\mathbf{2}$ & $1/2$ & $1$ \\
 $\Phi$ & 1 & $\mathbf{2}$ & $\overline{\mathbf{2}}$ & $0$ & $-1$ \\[1ex] 
 \hline
\end{tabular}
\vspace{3mm}
\caption{Leptonic and scalar particle content of the model.
\label{table:particleContent}}
\end{table}
\subsection{Yukawa interactions and scalar potential}
\label{subsec:yukpot}

The Yukawa Lagrangian of the model can be decomposed into a SM piece and a new piece involving the BSM fields,
\begin{align}  \label{eq:LY}
-\mathcal{L}_Y = -\mathcal{L}_Y^{\rm SM} + y_N \, \overline{L} \, \tilde{H} \, N +y_\chi\, \overline{L}\,\chi\,\tilde{\eta} + Y \,\overline{N^c} \operatorname{Tr}\left[\,\chi \, \Phi^\dagger\,\right]  + \frac{\mu}{2} \, \Tr{\left[\,\overline{\tilde{\chi}}\chi\,\right]} + \hc \, .
\end{align}
Here, $y_N$, $y_\chi$ and $Y$ are a $3 \times n_N$, $3 \times n_\chi$ and $n_N \times n_\chi$ general complex Yukawa matrices, while $\mu$ is a symmetric $n_\chi \times n_\chi$ complex matrix, with elements with dimensions of mass. The scalar potential can be written as
\begin{align}
\mathcal V &= m_H^2\, H^\dagger H +m_\eta^2\, \eta^\dagger \eta\,+ m_\Phi^2\, \operatorname{Tr}\left\{{\Phi^\dagger \Phi}\right\} \nonumber\\[3mm]
&+\frac{\lambda_H}{2}\left(H^\dagger H\right)^2 +\frac{\lambda_\eta}{2}\left(\eta^\dagger \eta\right)^2 + \frac{\lambda_\Phi}{2}\operatorname{Tr}\left\{\Phi^\dagger\Phi\right\}^2+ \frac{\lambda_\Phi'}{2}\operatorname{Tr}\left\{(\Phi^\dagger\Phi)^2\right\}\nonumber\\[3mm]&+\lambda_{H\eta}\,\left(H^\dagger H\right)\left(\eta^\dagger \eta\right)+ \lambda_{H\Phi}\,\left(H^\dagger H\right)\operatorname{Tr}\left\{{\Phi^\dagger \Phi}\right\}+\lambda_{\eta\Phi}\,\left(\eta^\dagger \eta\right)\operatorname{Tr}\left\{{\Phi^\dagger \Phi}\right\}\\[3mm]\nonumber
&+\kappa\left(H^\dagger\,\Phi\,\eta+\hc\right)\;.
\end{align}
The parameters $m_H^2$, $m_\eta^2$ and $m_\Phi^2$ have dimensions of mass$^2$ the cubic $\kappa$ has dimensions of mass while all quartic $\lambda_i$ couplings are dimensionless.

\subsubsection*{Symmetry breaking}

Given our set of scalars, the most general vacuum is given by
\begin{equation}
    \langle H\rangle= \frac{1}{\sqrt{2}}\left(\begin{array}{c}
         0  \\
         v_H 
         \end{array}\right)\;,\qquad
         \langle \eta\rangle= \frac{1}{\sqrt{2}}\left(\begin{array}{c}
         0  \\
         v_\eta 
         \end{array}\right)\;,\qquad \langle \Phi \rangle =\frac{1}{\sqrt{2}}\left(\begin{array}{cc}
        v_w & 0 \\
        0 & v_z
    \end{array}\right)\;.
\end{equation}
The minimization of the potential leads to four independent tadpole equations, which can be easily solved for the quadratic and cubic couplings,
\begin{align}
    m_H^2\,v_H +\frac{1}{\sqrt{2}}\kappa\, v_z\, v_\eta+ \frac{1}{2}\,\lambda_H\,v_H^3+\frac{1}{2}\lambda_{H\eta}\,v_\eta^2\,v_H+\frac{1}{2}\lambda_{H\Phi}\,(v_w^2+v_z^2)\,v_H&=0\;,\\[5mm]
    m_\eta^2\,v_\eta +\frac{1}{\sqrt{2}}\kappa\, v_z\, v_H+ \frac{1}{2}\,\lambda_\eta\,v_\eta^3+\frac{1}{2}\lambda_{H\eta}\,v_H^2\,v_\eta+\frac{1}{2}\lambda_{\eta\Phi}\,(v_w^2+v_z^2)\,v_\eta&=0\;,\\[5mm]
    m_\Phi^2\,v_w+\frac{1}{2}\lambda_\Phi\,(v_w^2+v_z^2)\,v_w+\frac{1}{2}\lambda_\Phi'\,v_w^3+\frac{1}{2}\,\lambda_{H\Phi}\,v_H^2\,v_w+\frac{1}{2}\lambda_{\eta\Phi}\,v_\eta^2\,v_w&=0\;, \label{eq:tad3} \\[5mm]
    \frac{1}{\sqrt{2}}\kappa\,v_H\,v_\eta+m_\Phi^2\,v_z+\frac{1}{2}\lambda_\Phi\,(v_w^2+v_z^2)\,v_z+\frac{1}{2}\lambda_\Phi'\,v_z^3+\frac{1}{2}\,\lambda_{H\Phi}\,v_H^2\,v_z+\frac{1}{2}\lambda_{\eta\Phi}\,v_\eta^2\,v_z&=0\;. \label{eq:tad4}
\end{align}
Some comments are in order regarding symmetry breaking. First, a non-vanishing vacuum expectation value (VEV) for $\Phi$ is required in order to break $SU(2)_1\times SU(2)_2 \times U(1)_Y$ down to the SM electroweak group, $SU(2)_L \times U(1)_Y$. Thus, we assume the following hierarchy
\begin{equation}
    v_\eta \ll  v_H \ll v_w \, , v_z \, .
\end{equation}
The condition $ v_H \ll v_w \, , v_z$ ensures that the SM is recovered at low energies, while $v_\eta \ll v_H$ guarantees that the Higgs VEV is the main source of electroweak symmetry breaking. As shown below in detail, this is an important phenomenological requirement in our model. Thus, SSB proceeds as
\begin{equation}
    SU(2)_1 \times SU(2)_2 \times U(1)_Y \times U(1)_{B-L} \xrightarrow{v_w \, , v_z} SU(2)_L \times U(1)_Y \xrightarrow{v_H \, , v_\eta} U(1)_{\text{em}}  \, ,
\end{equation}
where we have made use of the fact that $\langle \Phi \rangle$ also breaks spontaneously the $U(1)_{B-L}$ global group.

It is straightforward to show that an exact remnant $SU(2)_L$ symmetry is obtained only if $|v_w| = |v_z|$ (see Appendix~\ref{App:Symm} for more details). However, by inspecting Eqs.~\eqref{eq:tad3} and \eqref{eq:tad4} one can easily see that this equality can only be achieved if $\kappa = 0$. This choice introduces an accidental global $U(1)$ symmetry, giving rise to a doublet of NGBs that is phenomenologically excluded. Hence, the limit $v_w=v_z$ is experimentally ruled out. A phenomenologically viable realization of the model therefore requires a small departure from the exact $SU(2)_L$ limit, which we parametrize as $v_z = v_w + v_\Phi \, \delta$, with $v_\Phi \, \delta \ll v_w$. More specifically, we introduce the common bidoublet VEV $v_\Phi$ and impose
\begin{equation} \label{eq:vevhierarchy}
    v_{w,z} = v_\Phi \left( 1 \mp \frac{\delta}{2} \right) \, , \qquad \text{with} \qquad v_\eta \ll v_H \ll v_\Phi \, , \quad \delta \ll 1 \, .
\end{equation}
Here $v_\Phi \equiv \left( v_w + v_z \right)/2$ is the average bidoublet VEV and
\begin{equation} \label{eq:deltadef}
\delta \equiv \frac{v_z - v_w}{v_\Phi}
\end{equation}
 is a dimensionless measure of the splitting between the two. It is important to stress that $\delta$ is not an independent parameter. Combining the two tadpole equations of the bidoublet, Eqs.~\eqref{eq:tad3} and \eqref{eq:tad4}, the terms proportional to $m_\Phi^2$, $\lambda_\Phi$, $\lambda_{H\Phi}$ and $\lambda_{\eta\Phi}$ drop out and one is left with the exact relation
\begin{equation} \label{eq:deltatadpole}
\lambda_\Phi' \, v_z \left( v_z^2 - v_w^2 \right) = - \sqrt{2} \, \kappa \, v_H \, v_\eta
\qquad \Longrightarrow \qquad
\delta \simeq - \frac{\kappa \, v_H \, v_\eta}{\sqrt{2} \, \lambda_\Phi' \, v_\Phi^3} \, .
\end{equation}
The splitting of the bidoublet VEVs is therefore generated by the same coupling $\kappa$ that breaks the accidental symmetry discussed above and is proportional to $v_\eta$. Moreover, the tadpole equation for $\eta$ gives an induced VEV $v_\eta \simeq - \kappa \, v_H \, v_\Phi / (\sqrt{2} M_\eta^2)$, with $M_\eta^2 \equiv m_\eta^2 + \lambda_{\eta\Phi} v_\Phi^2$ the effective mass of the $\eta$ doublet. Combining both results,
\begin{equation} \label{eq:deltavalue}
\delta \simeq \frac{M_\eta^2 \, v_\eta^2}{\lambda_\Phi' \, v_\Phi^4} \, ,
\end{equation}
so that $\delta$ is a derived quantity, strongly suppressed for $v_\eta \ll v_H \ll v_\Phi$. For $v_\Phi = 10$ TeV, $M_\eta \sim 1$ TeV and $v_\eta \sim 1$ GeV one finds $\delta \sim 10^{-9}$. Together with
\begin{equation}
\varepsilon = v_H/v_\Phi \, ,
\end{equation}
$\delta$ is one of the two small parameters in which the particle spectrum is expanded. In what follows, the expressions for the particle spectrum are understood as expansions to leading order in both $\delta$ and $\epsilon$.
%

\section{Particle spectrum} \label{sec:Spectrum}

In this Section we analyze the particle spectrum of the model. We derive the masses after spontaneous symmetry breaking, paying particular attention to the features that distinguish the spectrum from that of more conventional scenarios. Unless otherwise stated, the symbol $\simeq$ denotes equalities that hold at leading order in the small parameters $\varepsilon$ and $\delta$.

\subsection{Gauge bosons}

The gauge sector of the model contains the kinetic terms
\begin{equation}\label{eq:kineticGaugeSector}
    \mathcal{L}_K=\left(D_\mu H\right)^\dagger\left(D^\mu H\right) +\left(D_\mu \eta\right)^\dagger\left(D^\mu \eta\right) + \operatorname{Tr}\left\{\left(D_\mu \Phi\right)^\dagger\left(D^\mu \Phi\right)\right\}\, ,
\end{equation}
where the covariant derivative is defined as usual,
\begin{equation} \label{eq:CovDerivative}
    D_\mu = \partial_\mu  -i  g_a\, A_\mu^a \, T^a   \, ,
\end{equation}
with $T^a$  the generators of the symmetry group.

The mass matrices for the neutral and charged gauge bosons can be found by evaluating Eq.~\eqref{eq:kineticGaugeSector} at the vacuum. Defining $V_n=\left(W_3^1 \ \ W_3^2 \ \ B\right)$ and $V_c=\left(W_+^1 \ \ W_+^2\right)$, we can write 
\begin{equation}    \mathcal{L}_m=V_n^\dagger\,\mathcal{M}^2_n\,V_n+V_c^\dagger\,\mathcal{M}^2_c\,V_c\;,\qquad
\end{equation}
with $\mathcal{M}_{n,c}$ the mass matrices for the neutral and charged fields, respectively. For the neutral gauge bosons, the mass matrix is given by
\begin{equation}
    \mathcal{M}_n^2=\frac{1}{4}\left(\begin{array}{ccc}
        g_1^2(v_H^2+v_w^2+v_z^2) & -g_1g_2(v_w^2+v_z^2) & -g_1g_Yv_H^2 \\
        -g_1g_2(v_w^2+v_z^2) & g_2^2(v_\eta^2+v_w^2+v_z^2) & -g_2g_Yv_\eta^2 \\
        -g_1g_Yv_H^2 & -g_2g_Yv_\eta^2 & g_Y^2(v_H^2+v_\eta^2)
    \end{array}\right)\;.
\end{equation}
From symmetry arguments, since the massless gauge field, the photon, must couple to the unbroken generator, $Q=T_3^1 + T_3^2 +Y$, we can infer
\begin{equation}
    A_\mu = \mathcal{N}_A\, \begin{pmatrix} g_2 g_Y & g_1 g_Y & g_1 g_2 \end{pmatrix} =\frac{1}{\sqrt{g_Y^2 + g^2}} \, \begin{pmatrix} g \,g_Y/g_1 & g\, g_Y/g & g \end{pmatrix}
\end{equation}
where we have defined the usual SM $SU(2)_L$ gauge coupling
\begin{equation}
   g \equiv \frac{g_1 \,g_2}{\sqrt{g_1^2 + g_2^2}} \, . 
\end{equation}
The massive states are given by some combination of vectors belonging to the 2-dimensional orthogonal space. Indeed, we can identify the combination
\begin{equation}
    W_L^3 = \frac{1}{\sqrt{g_1^2 + g_2^2}} \begin{pmatrix} g_2  & g_1 & 0 \end{pmatrix}
\end{equation}
playing the role of the $SU(2)_L$ neutral gauge boson in the SM. Therefore, just as in the SM, we expect the combination given by
\begin{align}
    Z_l = \frac{1}{\sqrt{g_Y^2 + g^2}} \begin{pmatrix} g  & -g_Y & 0 \end{pmatrix} \, ,
\end{align}
to play the role of the $Z$ boson in the SM. Thus, by defining the orthogonal,
\begin{align}
    Z_h = \frac{1}{\sqrt{g_Y^2 + g^2}} \begin{pmatrix} g_1  & -g_2 & 0 \end{pmatrix} \, ,
\end{align}
we find the following mass states,
\begin{align}
 \label{eq:Zmasses}
 Z &\simeq Z_l - \xi_Z \, Z_h \, , &&M_Z^2\simeq\frac{1}{4}(g^2+g_Y^2)v_H^2\;,\\
 Z' &\simeq Z_h + \xi_Z \, Z_l \, ,  && M_{Z'}^2\simeq\frac{1}{2}(g_1^2+g_2^2)v_\Phi^2\left(1+\frac{\delta^2}{4}\right)+\frac{1}{4}\frac{g^2g_1^2}{g_2^2}v_\Phi^2\,\varepsilon^2\;.
\end{align}
Two remarks are in order. First, $M_Z^2$ receives no correction in $\delta$: the leading one is of relative order $\delta^2 \varepsilon^2$, well beyond the accuracy retained here. Second, the factor $\left(1+\delta^2/4\right)$ in $M_{Z'}^2$ is exact, since the whole dependence on the splitting enters through the identity $v_w^2 + v_z^2 = 2 v_\Phi^2 \left(1 + \delta^2/4\right)$. We have also introduced the $Z$--$Z'$ mixing angle, which at leading order reads
\begin{equation} \label{eq:xiZ}
\xi_Z \simeq \frac{g^2}{2 \, g_2^2} \sqrt{\frac{g^2 + g_Y^2}{g_1^2 + g_2^2}} \; \varepsilon^2 \, .
\end{equation}

For the charged bosons, working in the $\left(W_+^1 \ \ W_+^2\right)$ basis we get:
%
\begin{equation}
    \mathcal{M}_c^2=\frac{1}{4}\left(\begin{array}{cc}
        g_1^2 \, (v_H^2+v_w^2+v_z^2) & -2g_1g_2v_w v_z \\
        -2g_1g_2 v_w v_z & g_2^2\, (v_\eta^2+v_w^2+v_z^2)
    \end{array}\right)\;.
\end{equation}
Again, we can identify the state playing the role of the $SU(2)_L$ boson and the orthogonal
\begin{align}
    W_l^+ &=  \frac{1}{\sqrt{g_1^2 + g_2^2}}\begin{pmatrix} g_2  & g_1  \end{pmatrix} \, , \\
    W_h^+ &=  \frac{1}{\sqrt{g_1^2 + g_2^2}}\begin{pmatrix} g_1  & -g_2  \end{pmatrix} \, ,
\end{align}
and then find the mass states,
\begin{align}
 \label{eq:Wmasses}
 W &\simeq W_l - \xi_W \, W_h \, , &&M_W^2\simeq\frac{1}{4}\, g^2 \left( v_H^2 + 2 v_\Phi^2 \delta^2 \right)\;,\\
 W' &\simeq W_h + \xi_W \, W_l  \, ,  && M_{W'}^2\simeq M_{Z'}^2-\frac{1}{2}\,g^2 v_\Phi^2 \, \delta^2\;. 
\end{align}
In contrast with $M_Z^2$, the $W$ mass does receive a correction proportional to $\delta^2$, the next one being of order $\delta^4$. This asymmetry between the two eigenstates is what generates a deviation of the $\rho$ parameter, as discussed in Sec.~\ref{subsec:ewpo}. The $W$--$W'$ mixing angle reads
\begin{equation} \label{eq:xiW}
\xi_W \simeq \frac{g^2 \, g_1}{2 \, g_2 \left( g_1^2 + g_2^2 \right)} \; \varepsilon^2 \, .
\end{equation}
The two angles are not independent. Indeed, one finds $\xi_W / \xi_Z = g/\sqrt{g^2+g_Y^2} = \cos\theta_W$, so that the neutral mixing is always the largest of the two.
Let us note that the charged and neutral heavy gauge bosons are degenerate up to terms of order $v_\Phi^2 \, \delta^2$, as expected since in the limit $v_H, v_\eta \to 0$ (hence $\delta \to 0$) they belong to the same triplet representation.


\subsection{Scalar bosons}

The model has four complex scalar fields, which after electroweak symmetry breaking give rise to a set of neutral and charged physical states. The corresponding mass matrices are lengthy, and we do not reproduce them here: they are collected in App.~\ref{App:Scalar}, where their diagonalisation is also discussed in detail. In this Section we restrict ourselves to the resulting mass eigenstates and eigenvalues, given at leading order in the small parameters of the model. We discuss the neutral and the charged sectors in turn.

We first decompose the neutral fields into their CP-even and CP-odd components,
\begin{align}
    H_0 &= \frac{1}{\sqrt{2}}(v_H+S_H+iA_H)\;, && 
    \eta_0 =\frac{1}{\sqrt{2}}(v_\eta+S_\eta+iA_\eta)\;,\nonumber\\[5mm]
    w_0&= \frac{1}{\sqrt{2}}(v_w+S_w+iA_w)\;,&&
    z_0= \frac{1}{\sqrt{2}}(v_z+S_z+iA_z)\;.\nonumber
\end{align}
We assume that the cubic coupling $\kappa$ in the scalar potential is real. Thus, CP is conserved (all the other couplings must be real for the Lagrangian to be Hermitian), and the CP-even and CP-odd states do not mix. This allows us to treat them separately.

In all sectors we prefer to work in the basis where the bidoublet components transform irreducibly under $SU(2)_L$, since in the limit in which this symmetry is conserved some general properties should be recovered. Using the decomposition $(\mathbf{2},\overline{\mathbf{2}})\to \mathbf{2}'\otimes\mathbf{2}' = \mathbf{3}'\oplus \mathbf{1}'$, where primes denote $SU(2)_L$ representations, the neutral components of the bidoublet with definite $SU(2)_L$ quantum numbers are the singlet $\phi_s$ and the neutral component of the triplet $\phi_t$,
\begin{align}
\phi_s = \frac{1}{\sqrt{2}} (w_0 + z_0) \;,&& \phi_t = \frac{1}{\sqrt{2}} (w_0 - z_0) \, ,
\label{eq:phist}
\end{align}
whose CP-even and CP-odd components we denote by
\begin{align} \label{eq:SAst}
S_{s,t} = \frac{1}{\sqrt{2}} \left( S_w \pm S_z \right) \;, && A_{s,t} = \frac{1}{\sqrt{2}} \left( A_w \pm A_z \right) \, .
\end{align}
Note that $\langle \phi_s \rangle = v_\Phi$ and $\langle \phi_t \rangle = - v_\Phi \, \delta /2$, so that the triplet VEV vanishes in the exact $SU(2)_L$ limit.

The two charged components of the bidoublet, $x_+$ and $y_-$, must first be brought to a common electric charge. Working with $x_+$ and $y^+ \equiv \left( y_- \right)^\dagger$, we define
\begin{equation} \label{eq:phistcharged}
\phi_s^+ = \frac{1}{\sqrt{2}} \left( x_+ + y^+ \right) \, , \qquad \phi_t^+ = \frac{1}{\sqrt{2}} \left( x_+ - y^+ \right) \, .
\end{equation}
In contrast with the neutral case, both combinations belong to the $SU(2)_L$ triplet: $\phi_s^+$ is the charged partner of $S_t$ in the CP-even real triplet, while $\phi_t^+$ is the charged partner of $A_t$ in the CP-odd one and is therefore eaten by the $W'$ boson.

To summarize, the CP-even, CP-odd and charged sectors are described in the bases
\begin{align} \label{eq:scalarbases}
\mathcal{S}^T &\equiv \left( S_H \ \ S_s \ \ S_t \ \ S_\eta \right) \, , &
\mathcal{A}^T &\equiv \left( A_H \ \ A_s \ \ A_t \ \ A_\eta \right) \, , &
\mathcal{S}^{+ \, T} &\equiv \left( H^+ \ \ \phi_s^+ \ \ \phi_t^+ \ \ \eta^+ \right) \, ,
\end{align}
respectively.

\subsubsection{Neutral scalars}

\paragraph{CP-odd scalars.}
%
For the CP-odd sector, as expected, we find three NGBs. Two of them, arising from the breaking of both $SU(2)$ symmetries, contribute to the longitudinal components of the $Z$ and $Z'$ gauge bosons, while the third one, the majoron, $J$, remains as a physical NGB associated with the spontaneous breaking of $U(1)_{B-L}$. In the basis $\mathcal{A}$ of Eq.~\eqref{eq:scalarbases}, the physical states can be written as
\begin{align}
    \label{eq:P1mass}
    \P_1 &\simeq \mathcal{N}_{\P_1} \left(- \frac{v_\eta}{v_H} \ \ \frac{v_\eta}{\sqrt{2}v_\Phi} \ \ \frac{-v_\eta}{\sqrt{2}v_\Phi} \ \ 1 \right) \; , && M_{\P_1}^2 = \lambda_\Phi' \frac{v_\Phi^4}{v_\eta^2} \, \delta = M_\eta^2  \, ,\\
   J &\simeq \mathcal{N}_J \left( \frac{v_\eta^2}{\sqrt{2} \, v_H v_\Phi} \ \ 1 \ \ \frac{\delta}{2} \ \ \frac{-v_\eta}{\sqrt{2} v_\Phi} \right) \; , &&  M_{J}^2 = 0 \label{eq:Jprofile}
\end{align}
where $\mathcal{N}_{\P_1}$ and $\mathcal{N}_J$ are two normalization factors. We note that $\P_1$ becomes massless in the limit $\delta \to 0$. As already pointed out in Sec.~\ref{subsec:yukpot}, this requires $\kappa \to 0$, which in turn leads to an accidental $U(1)$ symmetry that is spontaneously broken by the $\Phi$ VEV. Hence, $\P_1$ would be the NGB associated to this accidental symmetry, explicitly broken by $\kappa \neq 0$. We refer to App.~\ref{App:Scalar} for more details on the mass matrix and its diagonalization.

\paragraph{CP-even scalars.}
As anticipated above, we work in the basis of Eq.~\eqref{eq:SAst}, in which the bidoublet components transform irreducibly under $SU(2)_L$.
In the basis $\mathcal{S}$ of Eq.~\eqref{eq:scalarbases}, the physical states and masses are given by
\begin{align}
\S_1 &\simeq 
\begin{pmatrix}
   1 &
\frac{\sqrt{2} v_H \lambda_{H\Phi}}{v_\Phi\left(2\lambda_\Phi+\lambda_\Phi'\right)}
&
\frac{\sqrt{2}\,v_\Phi \delta}{v_H}
&
-\frac{v_\eta}{v_H}
\end{pmatrix}
\, , && M_{\S_1}^2 \simeq  \lambda_H v_H^2 \left(1  - \frac{2 \lambda_{H\Phi}^2}{\lambda_H \left( 2 \lambda_\Phi + \lambda_\Phi' \right)} \right) \label{eq:S1mass}  \,  
\\[2ex]
\S_2 &\simeq
\begin{pmatrix}
  \frac{\sqrt{2} v_H \lambda_{H\Phi}}{v_\Phi\left(2\lambda_\Phi+\lambda_\Phi'\right)}
&
1
&
0
&
-\frac{v_\eta}{\sqrt{2} v_\Phi}  
\end{pmatrix}
\,, && M_{\S_2}^2 \simeq \left(2 \lambda_\Phi+\lambda_\Phi'\right) v_\Phi^2 
\\[2ex]
\S_3 &\simeq
\begin{pmatrix}
-\frac{\sqrt{2}\,v_\Phi \delta}{v_H}
&
0
&
1
&
\frac{v_\eta}{\sqrt{2} v_\Phi}
\end{pmatrix}
\,, && M_{\S_3}^2 \simeq \lambda_\Phi' v_\Phi^2  
\\[2ex]
\S_4 &\simeq
\begin{pmatrix}
\frac{v_\eta}{v_H}
&
\frac{v_\eta}{\sqrt{2} v_\Phi}\left(1+ \frac{\lambda_\Phi'}{2\lambda_\Phi+\lambda_\Phi'}\right)
&
-\frac{v_\eta}{\sqrt{2} v_\Phi}
&
1
\end{pmatrix}
\,, && M_{\S_4}^2 \simeq \lambda_\Phi' \frac{v_\Phi^4}{v_\eta^2} \delta = \, M_\eta^2 \, , \label{eq:S4mass}
\end{align}
Let us note that $\S_1$ corresponds to the SM-like Higgs. Again, we refer to App. \ref{App:Scalar} for more details.

\subsubsection{Charged scalars}

The charged scalar sector involves the charged components of the four multiplets, in the basis $\mathcal{S}^{+}$ of Eq.~\eqref{eq:scalarbases}. The mass matrix $\M_{S^+}^2$, given explicitly in App.~\ref{subsec:charged}, contains two vanishing eigenvalues corresponding to the NGBs that become the longitudinal components of $W$ and $W'$. For the physical states we obtain
%
\begin{align}
    \mathcal{H}_1^+ &\simeq \begin{pmatrix}
        \frac{-v_\eta}{\sqrt{2}v_\Phi} & \frac{v_\eta}{\sqrt{2}v_\Phi} & \frac{-v_\eta }{v_H} & 1 \end{pmatrix} \; , && M_{\mathcal{H}_1^+}^2 \simeq \lambda_\Phi' \frac{v_\Phi^4}{v_\eta^2} \delta = \, M_\eta^2 \, , \label{eq:H1mass} \\
      \mathcal{H}_2^+ &\simeq \begin{pmatrix}
        \frac{-\sqrt{2}v_\Phi \delta}{v_H} & 1 & \delta & \frac{ v_\eta }{\sqrt{2} v_\Phi}  \end{pmatrix}\; , && M_{\mathcal{H}_2^+}^2 \simeq \lambda_\Phi' v_\Phi^2  \, . 
\end{align}
Again, more details can be found in App.~\ref{App:Scalar}.

\subsection{Fermions}

The quark sector of the model is identical to that of the SM, since no new colored states are introduced and the quarks are singlets under $SU(2)_2$. We therefore restrict our discussion to the lepton sector, which is where all the new fields reside.
We follow the same conventions as in the scalar sector. The interaction
eigenstates are collected in the vectors $\mathcal{F}^0$ and $\mathcal{F}^\pm$,
and the corresponding mass eigenstates are denoted by $\mathcal{F}^0_i$, with
$i = 1 \ldots 3 + 3 n_F$, and $\mathcal{F}^\pm_i$, with $i = 1 \ldots 3 + n_F$,
ordered by increasing mass. The three lightest neutral states,
$\mathcal{F}^0_{1,2,3}$, are the light active neutrinos, and the three lightest
charged ones, $\mathcal{F}^\pm_{1,2,3}$, are the SM charged leptons. The
remaining $3 n_F$ neutral and $n_F$ charged states are the new heavy leptons.
When needed, we collect them in the vectors
$\mathcal{F}^0_\text{m} \equiv \left( \mathcal{F}^0_1 \ \ldots \ \mathcal{F}^0_{3 + 3 n_F} \right)^T$
and $\mathcal{F}^\pm_\text{m} \equiv \left( \mathcal{F}^\pm_1 \ \ldots \ \mathcal{F}^\pm_{3 + n_F} \right)^T$.\\

Before discussing each sector separately, it is convenient to introduce the following decomposition into states with a definite transformation under $SU(2)_L$, just as we did in the scalar sector. Using that for bidoublets we have
\begin{align}
 \chi = \frac{1}{\sqrt{2}} \left( \chi_s \, \mathbf{1} + \chi_t^a \, \sigma^a\right)  \, , &&   \Phi = \frac{1}{\sqrt{2}} \left( \phi_s \, \mathbf{1} + \phi_t^a \, \sigma^a\right) \, ,
\end{align}
with $\sigma^a$ the Pauli matrices, the leptonic Yukawa sector can be written as
\begin{align} \label{eq:Yukawadecom}
-\mathcal{L} \subset &y_e \, \overline{L} \, H \, e_R+ y_N \, \overline{L} \, \tilde{H} \, N +y_\chi\, \overline{L}\,\chi\,\tilde{\eta} + Y \,\overline{N^c} \operatorname{Tr}\left[\,\chi \, \Phi^\dagger\,\right]  + \frac{\mu}{2} \, \Tr{\left[\,\overline{\tilde{\chi}}\chi\,\right]} + \hc \nonumber 
\\[0.25cm]
= \; \; &\,y_e \, \overline{L} \, H \, e_R+ y_N \, \overline{L} \, \tilde{H} \, N +\frac{y_\chi}{\sqrt{2}}\, \overline{L}\,\tilde{\eta}\,\chi_s + Y \,\phi_s^\dagger \, \overline{N^c} \,\chi_s + \frac{\mu}{2} \,\overline{\tilde{\chi}}_s\,\chi_s \nonumber 
\\
+&\frac{y_\chi}{\sqrt{2}}\, \overline{L}\,\tilde{\eta}\,\sigma^a \, \chi_t^a + Y \, \overline{N^c} \,\chi_t^a \, \left(\phi_t^a\right)^\dagger + \frac{\mu}{2} \,\overline{\tilde{\chi}}_t^{\, a} \,\chi_t^a \, +\hc\;.
\end{align}
With this decomposition now it is trivial to identify the role played by each fermion.

\subsubsection{Charged fermions} \label{subsubsec:chargedfermions}

The relevant terms in the charged lepton sector are
\begin{align} 
-\mathcal{L} \subset &\, y_e \, \overline{L} \, H \, e_R +  \frac{y_\chi}{\sqrt{2}}\, \overline{L}\,\tilde{\eta}\,\sigma^a \, \chi_t^a  + \frac{\mu}{2} \,\overline{\tilde{\chi}}_t^{\, a} \,\chi_t^a \, .
\end{align}
This Lagrangian is equivalent to that of a Dirac Type-III seesaw, where $\chi_t$ plays the role of the heavy fermion. In the basis $\mathcal{F}^\pm_{L,R} \equiv \left(e \ \ \chi^- \right)_{L,R}$, the mass term can be written as
\begin{align}
-\mathcal{L}_\pm^m=\overline{\mathcal{F}}^\pm_L\,\mathcal{M}_\pm\,\mathcal{F}^\pm_R+\hc\, ,
\end{align}
with the mass matrix
\begin{equation}
    \qquad\mathcal{M}_\pm=\left(\begin{array}{cc}
        M_\ell & \frac{v_\eta\,y_\chi}{\sqrt{2}} \\
         0 & \mu
    \end{array}\right)\, .
\end{equation}
Here we have introduced the SM charged lepton mass matrix
\begin{equation}
M_\ell = \frac{v_H\,y_e}{\sqrt{2}} \, .
\end{equation}
In order to obtain SM-like charged leptons, we must require either $\mu \gg M_\ell$ or $\mu \ll M_\ell$. Since the latter would result in light charged fermions, which are phenomenologically excluded, we focus on the former possibility, $\mu \gg M_\ell$. In this regime, we can use the seesaw expansion described in App.~\ref{App:Seesaw} to obtain
\begin{align}
    \mathcal{M}^{\text{m}}_\pm \simeq \begin{pmatrix}
        M_\ell & 0 \\
        0 & \mu
    \end{pmatrix} \, , 
    &&
   \mathcal{F}^\pm_{L,R} = \begin{pmatrix}
       \sqrt{\id_3 -  B_{L,R}B_{L,R}^\dagger} & B_{L,R}\\
       -B_{L,R}^\dagger & \sqrt{1-  B_{L,R}^\dagger B_{L,R}}
   \end{pmatrix} \, \mathcal{F}^\pm_{\text{m} \, L,R} \,
\end{align}
with
\begin{align}
    B_L \simeq \frac{v_\eta}{\sqrt{2}}  y_\chi \mu^{-1} \; , && B_R \simeq \frac{v_\eta}{\sqrt{2}} M_\ell \,  y_\chi  \mu^{-1}\, ,
\end{align}
and the superscript/subscript m refers to the mass basis.
\subsubsection{Neutral fermions}\label{subsubsec:neutralfermions}
In this case we will focus on the following Lagrangian,
\begin{align}
-\mathcal{L} \supset
\; &\linInv{y_N \, \overline{L} \, \tilde{H} \, N}
\;+\; \linear{\frac{y_\chi}{\sqrt{2}}\, \overline{L}\,\tilde{\eta}\,\chi_s}
\;+\; \linInv{Y \,\phi_s^\dagger \, \overline{N^c} \,\chi_s}
\;+\; \inverse{\frac{\mu}{2} \,\overline{\tilde{\chi}}_s\,\chi_s}
\nonumber \\[10pt]
+\; &\typeI{\frac{y_\chi}{\sqrt{2}}\, \overline{L}\,\tilde{\eta}\,\sigma^a \, \chi_t^a}
\;+\; Y \, \overline{N^c} \,\chi_t^a \, \left(\phi_t^a\right)^\dagger
\;+\; \typeI{\frac{\mu}{2} \,\overline{\tilde{\chi}}_t^{\, a} \,\chi_t^a} \, .
\end{align}
Here, we use colors to identify the different mass-generation mechanisms arising in our model, with blue, red, and green denoting the type-III, inverse and linear seesaw mechanisms, respectively. Thus, one expects the model to realize a hybrid seesaw mechanism. We note that the term involving $\phi_t^a$ contributes to the mass matrix, but is proportional to $\delta$ and therefore gives rise to a subdominant contribution. Indeed, considering the full Lagrangian and defining $\mathcal{F}^0 \equiv \left(\nu_L^c \ \ N_R \ \ \chi_s \ \ \chi_t \right)$, we find
\begin{align}
    -\mathcal{L}_0^m=\frac{1}{2}\,\overline{(\mathcal{F}^0)^c}\,\mathcal{M}_0\,\mathcal{F}^0+\hc\,,
\end{align}
with
\begin{align}
    \mathcal{M}_0  \equiv 
    \begin{pmatrix}
        0     & M_D   \\
        M_D^T & M  
    \end{pmatrix} \, ,
\end{align}
where we have introduced the following matrices
\begin{align}
 M_D\equiv 
   \frac{1}{\sqrt{2}} \begin{pmatrix} v_H y_N & v_\eta y_\chi/\sqrt{2} & v_\eta y_\chi/\sqrt{2}      \end{pmatrix} \; , \; \; M \equiv
   \begin{pmatrix}
     0 & Y\,\frac{2 v_\Phi}{2 \sqrt{2}} & Y\,\frac{v_w-v_z}{2\sqrt{2}} \\
     Y^T\,\frac{2 v_\Phi}{2\sqrt{2}}  & -\mu & 0 \\
     Y^T\,\frac{v_w-v_z}{2\sqrt{2}} & 0 & \mu \\ 
   \end{pmatrix} \, .
\end{align}
Again, we can use the seesaw expansion introduced in App.~\ref{App:Seesaw}. Thus, the mass states are given by
\begin{align}
    \mathcal{F}^0 &= \begin{pmatrix}
       \sqrt{\id_3 - P P^\dagger} & P\\
       -P^\dagger & \sqrt{\id_6-  P^\dagger P}
   \end{pmatrix} \, 
   \begin{pmatrix}
    U_{\text{PMNS}} & 0 \\
    0 & U_H
   \end{pmatrix} \, \mathcal{F}^0_{\text{m}}  \nonumber
   \\[0.5cm]
   &\simeq \begin{pmatrix}
       \id_3 - \frac{1}{2} P P^\dagger & P\\
       -P^\dagger & \id_6- \frac{1}{2} P^\dagger P
   \end{pmatrix} \, 
   \begin{pmatrix}
    U_{\text{PMNS}} & 0 \\
    0 & U_H
   \end{pmatrix} \, \mathcal{F}^0_{\text{m}} \, ,
\end{align}
with
\begin{align}
    P \simeq M_D^* \left( M^{-1} \right)^\dagger \, , \label{eq:Pneutral}
\end{align}
the matrix that parametrizes the mixing between light and heavy states. $ U_{\text{PMNS}}$ and $U_H$ are the unitary matrices that diagonalize the light and heavy mass matrices, $M_\nu$ and $M_H$, respectively, which are given at leading order by
\begin{align}
    M_\nu &\simeq
    \cunderbrace{seeI}{- \frac{v_\eta^2}{4} \, y_\chi \, \mu^{-1} \, y_\chi^T}{type-III}
    \; \cunderbrace{seeInv}{- \frac{v_H^2}{v_\Phi^2} \, y_N \, \left(Y^T \right)^{-1} \, \mu \, Y^{-1} \,y_N^T}{inverse}
    \; \cunderbrace{seeL}{- \frac{v_H \, v_\eta}{2 \, v_\Phi} \left( y_N \left(Y^T \right)^{-1} y_\chi^T + y_\chi Y^{-1} y_N^T \right)}{linear} \, , 
    \\
    M_H &\simeq M \, .
\end{align}
Let us analyze the hybrid structure related to  light neutrino masses. First, it is straightforward to see that with $n_N = n_\chi \equiv n_F$, the matrix $Y$ is square and the light neutrino mass matrix takes the factorized form $M_\nu = - W W^T$, with
\begin{equation} \label{eq:W}
W \equiv \frac{v_\eta}{2} \, y_\chi \, \mu^{-1/2}
       + \frac{2 v_H}{v_z + v_w} \, y_N \left( Y^T \right)^{-1} \mu^{1/2}
\end{equation}
a $3 \times n_F$ matrix. Therefore
\begin{equation}
\text{rank} \left( M_\nu \right) = \text{rank} \left( W \right) = \min \left( 3, n_F \right) \, ,
\end{equation}
and the number of massive light neutrinos is simply given by the number of heavy
lepton families. Since oscillation experiments have measured two independent
mass-squared differences, at least two light neutrinos must be massive, and hence
$n_F \geq 2$. In what follows we adopt the minimal choice $n_F = 2$, which
predicts a massless lightest neutrino.

\begin{figure}[t]
    \centering
    \includegraphics[width=0.72\textwidth]{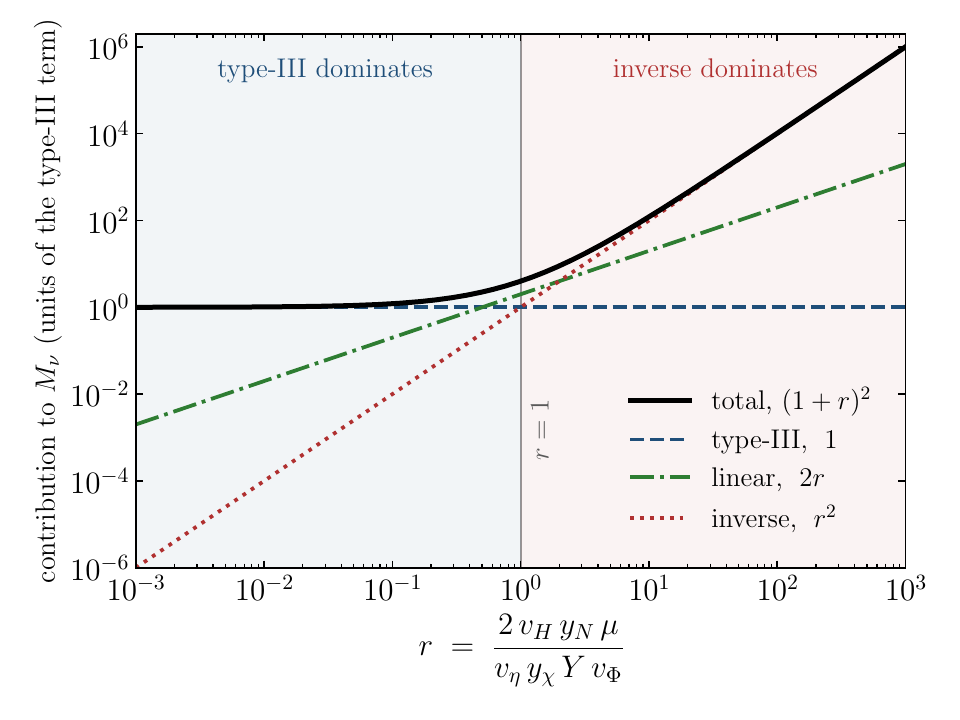}
    \caption{The three contributions to the light neutrino mass matrix, normalised
    to the type-III one, as a function of the ratio $r$ defined in
    Eq.~\eqref{eq:ratio}. The linear contribution never dominates.}
    \label{fig:seesaw_r}
\end{figure}

\begin{figure}[t]
    \centering
    \includegraphics[width=0.75\textwidth]{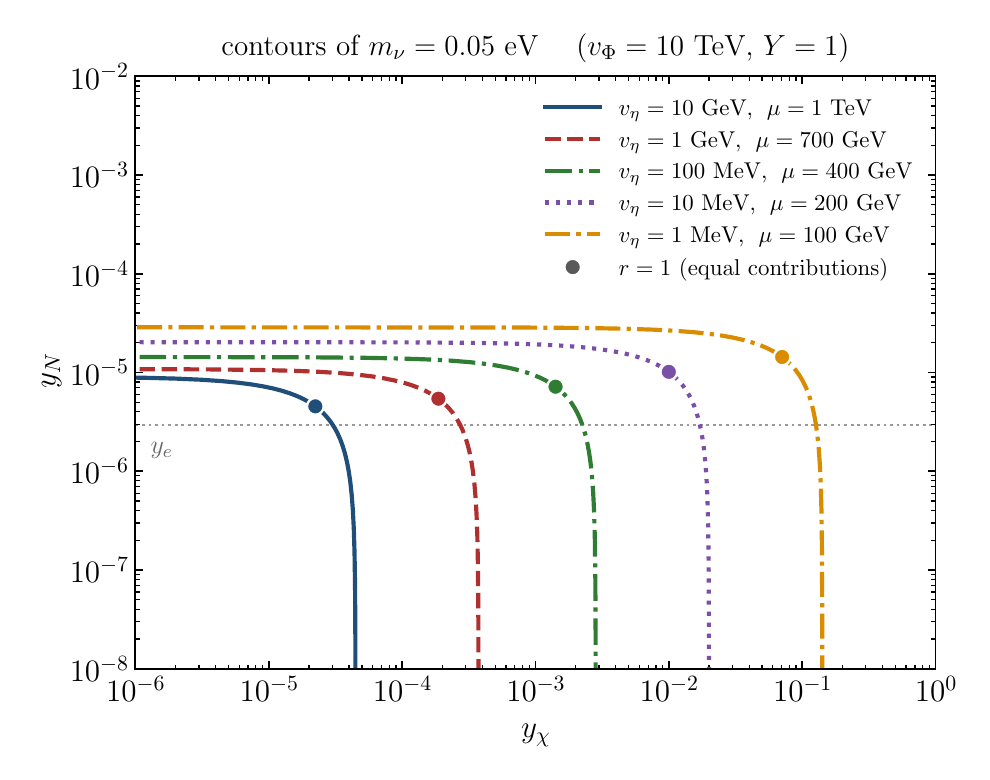}
    \caption{Contours reproducing $m_\nu = 0.05$ eV in the $(y_\chi, y_N)$ plane,
    for several values of $v_\eta$ and $\mu$, with $v_\Phi = 10$ TeV and $Y = 1$.
    The dots mark the points where the type-III and inverse contributions are equal,
    $r = 1$. The dotted horizontal line
    indicates the electron Yukawa coupling, $y_e = \sqrt{2} \, m_e / v_H$, for
    reference. A single generation is assumed, so that the Yukawa couplings are
    numbers.}
    \label{fig:mnu_yukawa}
\end{figure}

The factorized form $M_\nu = - W W^T$ also allows one to understand which of the
three seesaw contributions dominates. Expanding $W W^T$, the type-III, linear and
inverse terms appear in the ratio
\begin{equation} \label{eq:ratio}
\text{type-III} \, : \, \text{linear} \, : \, \text{inverse} \; = \; 1 \, : \, 2r \, : \, r^2 \, ,
\qquad
r \equiv \frac{2 \, v_H \, y_N \, \mu}{v_\eta \, y_\chi \, Y \, v_\Phi} \, ,
\end{equation}
so that a single combination of parameters controls the hierarchy among them. This
is illustrated in Fig.~\ref{fig:seesaw_r}. The type-III contribution dominates for
$r \ll 1$ and the inverse one for $r \gg 1$, while the linear term never dominates
on its own: it is largest, relative to the others, precisely in the transition
region $r \simeq 1$, where all three are comparable.

An immediate consequence of Eq.~\eqref{eq:ratio} is that $M_\nu \propto (1+r)^2$,
so that the light neutrino masses would vanish for $r = -1$. This is a genuine
cancellation between the two terms in $W$, and it would allow one to reproduce the
observed neutrino masses with sizable Yukawa couplings, and hence with potentially
observable lepton flavor violation. We stress, however, that this alignment is not
enforced by any symmetry of the model, and we will not pursue it further.\\

On the other hand, it is important to note that the interplay between the different contributions to the light neutrino mass matrix makes its structure more complex than it might seem at first sight. At first glance, one might conclude that the inverse and linear seesaw mechanisms allow the model to realize a low-scale seesaw, with all its associated features. However, this is not the case. Indeed, the mixing between light and heavy neutral fermions is given by Eq.~\eqref{eq:Pneutral}, and scales as
\begin{align}
    P \sim \sqrt{\frac{m_\nu}{\mu}} \, .
\end{align}
This suppression is typical of the inverse seesaw and would therefore usually be associated with a low-scale seesaw scenario. However, in the present model, it also suppresses the type-III seesaw contribution. Consequently, the light–heavy neutrino mixing remains suppressed by the light-neutrino mass, as in the conventional type-III seesaw, preventing the model from exhibiting the enhanced mixing characteristic of low-scale seesaw scenarios. This can be also understood as follows. The parameter $\mu$, which conventionally sets the low scale in the inverse seesaw, also sets the mass scale of the type-III seesaw contribution in our model. Therefore, taking $\mu$ to be small, as traditionally done in an inverse seesaw scenario, would lead to an excluded scenario for two reasons: first, it would yield unrealistic neutrino masses, and second, it would give rise to a light fermion triplet, and hence to new light charged fermions. Moreover, the fact that $\mu$ sets the mass scale of the new charged fermions imposes a lower bound on its value, which we take to be $\mu \gtrsim 0.1-1$ TeV~\cite{ATLAS:2022yhd}.~\footnote{We note that conventional searches for charged fermions in the type-III seesaw assume the canonical type-III seesaw framework. The more complex structure of our model may therefore modify the corresponding analyses and, consequently, the resulting experimental limits. We thus adopt an approximate constraint, while a dedicated analysis would be required to establish the precise experimental limits for the specific realization of the type-III charged fermions considered here.} Then, from the inverse seesaw contribution, one can obtain
\begin{align} \label{eq:vphiboundinverse}
    m_\nu \gtrsim \frac{y_N^2}{Y^2}\, \frac{v_H^2}{v_\Phi^2} \, \mu \; \Rightarrow \quad v_\Phi \gtrsim \frac{y_N}{Y}\, v_H \,\sqrt{\frac{\mu}{m_\nu}} \,.
\end{align}
For instance, taking $m_\nu \sim 0.05$ eV we find
\begin{align} \label{eq:vphimnubound}
    v_\Phi \gtrsim \frac{y_N}{Y}\,\sqrt{\frac{\mu}{\textup{TeV}}} \, 10^{6} \; \textup{TeV}\gtrsim \frac{y_N}{Y}\,10^{5-6} \; \textup{TeV} \,,
\end{align}
where in the last step we have used $\mu \gtrsim 0.1-1$ TeV. Thus, one needs very high scales or very small $y_N$, both departing from the low-scale regime. This is illustrated in Fig.~\ref{fig:mnu_yukawa}, where we display the contours reproducing $m_\nu = 0.05$ eV in the $(y_\chi, y_N)$ plane for $v_\Phi = 10$ TeV and $\mu$ in the range of hundreds of GeV. Each contour has the shape of an L: along the horizontal branch the neutrino mass is controlled by $y_N$ through the inverse term, while along the vertical branch it is controlled by $y_\chi$ through the type-III one, the corner corresponding to $r = 1$. One can see that explaining neutrino masses for sizable values of $v_\Phi$ requires very small values of $y_N$, namely, $y_N \lesssim 10^{-(4-5)}$. In the region where the type-III contribution dominates, however, only the combination $v_\eta\, y_\chi$ needs to be sufficiently suppressed, as can be inferred from the different lines.  

\subsection{Summary of the spectrum}

Having discussed the particle spectrum of the model in detail, let us summarize it. \\

\textbf{Gauge bosons: SM + $Z'$, $W'$.} In addition to the usual SM gauge bosons,
the model includes heavy $Z'$ and $W'$ states, associated to the breaking
$SU(2)_1 \times SU(2)_2 \to SU(2)_L$, with masses of the order of $v_\Phi$. They
are degenerate up to corrections of order $\delta^2$, a consequence of the residual $SU(2)$ symmetry, and mix with their SM counterparts with angles suppressed by $\varepsilon^2$. \\

\textbf{Scalars: $\S_{1,2,3,4}$, $\P_1$, $J$, $\mathcal{H}_{1,2}^+$.} The neutral
CP-even sector contains four physical states. The lightest one is a SM-like Higgs
boson, $\S_1 \equiv h$, while $\S_2$ and $\S_3$ have masses of the order of
$v_\Phi$. In the CP-odd sector, three of the four states are NGBs: two of them are eaten by the $Z$ and $Z'$ bosons, while the third one, the majoron $J$, remains in the spectrum as a physical massless state associated to the spontaneous breaking of $U(1)_{B-L}$. The remaining CP-odd state is $\P_1$. Finally, two physical singly charged scalars, $\mathcal{H}_1^+$ and $\mathcal{H}_2^+$, are left after the NGBs eaten by $W$ and $W'$ are removed. We note that $\P_1$, $\S_4$ and $\mathcal{H}_1^+$ are degenerate, with a common mass given by the effective mass of the $\eta$ doublet. \\

\textbf{Leptons: $\mathcal{F}^0_{1 \ldots 3 + 3 n_F}$, $\mathcal{F}^\pm_{1 \ldots 3 + n_F}$.} In addition to the SM leptons, the
model contains $n_F$ generations of the singlet fermions $N$ and of the bidoublets $\chi$, which decompose into $SU(2)_L$ singlets and triplets. In the neutral sector, the light active neutrinos acquire masses through a hybrid of the type-III, inverse and linear seesaw mechanisms, while the heavy states have masses controlled by $\mu$ and by $Y v_\Phi$. Since the light neutrino mass matrix takes the factorized form $M_\nu = - W W^T$, its rank equals $n_F$, so that the minimal choice $n_F = 2$ gives two massive neutrinos and leaves the lightest one massless. In the charged sector, the SM charged leptons mix with the charged components of $\chi$, with mixing angles suppressed by $v_\eta / \mu$. \\

\begin{figure}[t]
    \centering
    \includegraphics[width=0.8\textwidth]{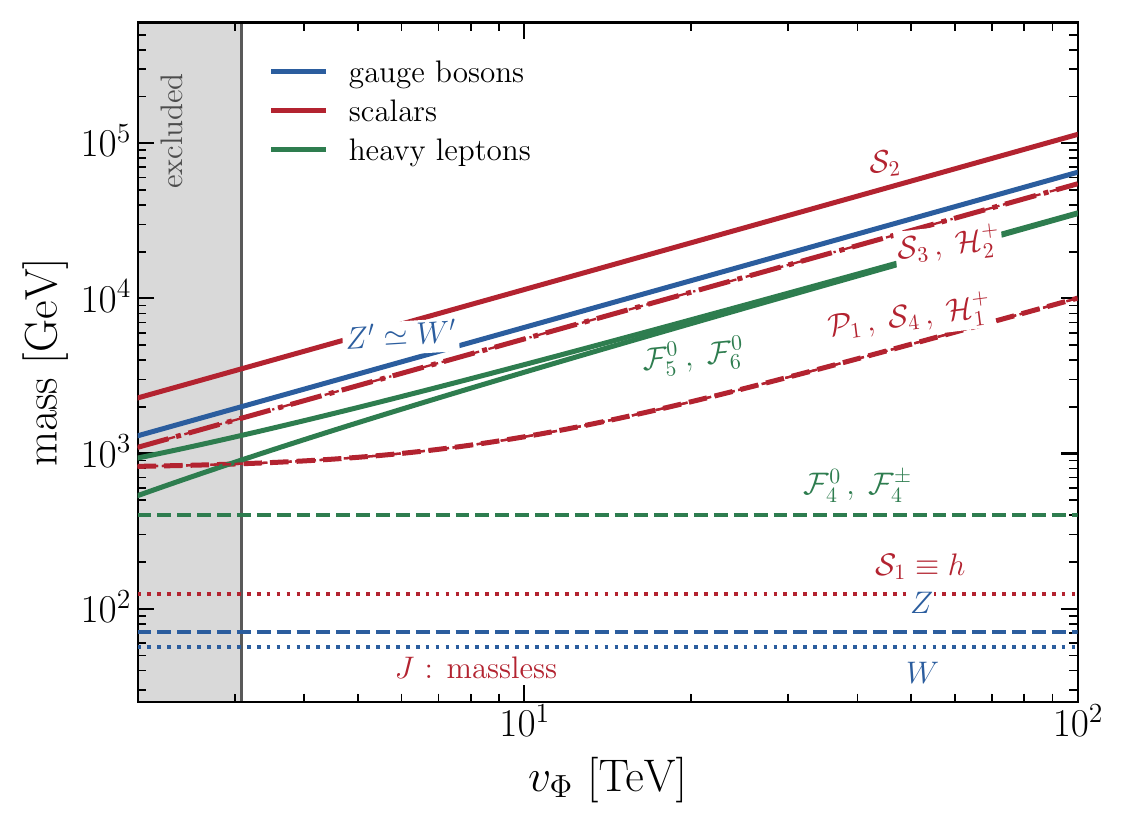}
    \caption{Mass spectrum of the model as a function of $v_\Phi$. We have fixed
    $g_1 = g_2 = 0.65$, $\lambda_\Phi = 0.5$, $\lambda_\Phi' = 0.3$,
    $\lambda_{H\Phi} = 0.05$, $\lambda_{H\eta} = 0.1$, $\lambda_{\eta\Phi} = 0.01$, $\lambda_\eta = 0.5$, $m_\eta = 800$ GeV, $\kappa = -0.1$ GeV, $\mu = 400$ GeV and $Y = 0.5$, while $\lambda_H$ is adjusted to reproduce $m_h = 125$ GeV. At each point, $v_\eta$ and $\delta$ are obtained by solving the tadpole equations. The grey region is excluded by gauge boson mixing, see Sec.~\ref{subsec:ewpo}.}
    \label{fig:spectrum}
\end{figure}

The resulting spectrum is illustrated in Fig.~\ref{fig:spectrum}, where we show the masses of all the states of the model as a function of the symmetry breaking scale $v_\Phi$, for a representative choice of the remaining parameters. Several features are worth noticing. The heavy gauge bosons are degenerate, $M_{W'} \simeq M_{Z'}$, as are the scalars $\mathcal{S}_3$ and $\mathcal{H}_2^+$, which are the neutral and charged members of the same $SU(2)_L$ triplet, both degeneracies being a consequence of the residual $SU(2)$ symmetry. The states $\mathcal{P}_1$, $\mathcal{S}_4$ and $\mathcal{H}_1^+$ are also degenerate, but for a different reason: they are dominantly the components of the quasi-inert doublet $\eta$, and their common mass approaches $m_\eta$ at low $v_\Phi$ and grows as $\sqrt{\lambda_{\eta\Phi}} \, v_\Phi$ once the second term dominates. The majoron $J$ is exactly massless and the lightest CP-even state is the SM-like
Higgs boson. Finally, the gray region on the left is excluded by the gauge boson
mixing constraints that we will derive in Sec.~\ref{subsec:ewpo}.

\section{Phenomenology} \label{sec:Pheno}

The particle spectrum derived in Sec.~\ref{sec:Spectrum} is controlled, to a large extent, by only two quantities: the bidoublet VEV $v_\Phi$, which sets the scale of the $SU(2)_1 \times SU(2)_2 \to SU(2)_L$ breaking, and the splitting $v_\Phi \, \delta$, which measures the departure from an exact $SU(2)_L$. This is the origin of the correlations discussed in this Section. We first discuss electroweak precision constraints, then collider signatures, and finally the LFV and majoron phenomenology.

\subsection{Electroweak precision constraints} \label{subsec:ewpo}

\subsubsection*{Gauge boson mixing}

Gauge boson mixing is known to lead to stringent constraints in models with extended gauge groups, in particular in \3221~\cite{Boucenna:2016wpr,Boucenna:2016qad}. In our model, the new gauge bosons mix with the SM ones already at tree level, with the mixing angles $\xi_Z$ and $\xi_W$ given in Eqs.~\eqref{eq:xiZ} and \eqref{eq:xiW}. Both are suppressed by $\varepsilon^2 = v_H^2/v_\Phi^2$ and, crucially, are independent of $\delta$. Moreover, they are not independent of each other, since $\xi_W/\xi_Z = \cos\theta_W$, so that it suffices to constrain the neutral mixing.

Two features are worth emphasizing. First, the mixing is controlled by the ratio of gauge couplings: for $g_2 \gg g_1$ the angles scale as $\xi_Z \sim g_1^2/g_2^2$ and the mixing is naturally suppressed, whereas for $g_1 \sim g_2$ it is of order $\varepsilon^2$. Second, and in contrast with the model of Ref.~\cite{Boucenna:2016wpr}, the mixing cannot be switched off by adjusting the ratio of the doublet VEVs: here the hierarchy $v_\eta \ll v_H$ is dictated by the neutrino mass mechanism and cannot be tuned.

Rather than performing a global fit to $Z$- and $W$-pole pseudo-observables, as was done in Refs.~\cite{Boucenna:2016wpr,Boucenna:2016qad}, we simply require the mixing angle to lie below the level typically allowed by electroweak precision analyses of models with an extra neutral gauge boson, $|\xi_Z| \lesssim \xi_\text{max}$. Using Eq.~\eqref{eq:xiZ} this translates into a lower bound on the symmetry breaking scale,
\begin{equation} \label{eq:vPhibound}
v_\Phi \gtrsim v_H \left[ \frac{g^2}{2 \, g_2^2 \, \xi_\text{max}} \sqrt{\frac{g^2+g_Y^2}{g_1^2+g_2^2}} \right]^{1/2} \, .
\end{equation}
For $g_1 = g_2$ and $\xi_\text{max} = 10^{-3}$ this gives $v_\Phi \gtrsim 3$ TeV, while for $g_2 \gg g_1$ the bound is considerably weaker. We stress that $\xi_\text{max} = 10^{-3}$ is only indicative. Limits on the $Z$--$Z'$ mixing of this order are common in the literature~\cite{Hsieh:2010zr,Boucenna:2016wpr,Boucenna:2016qad}, but they depend strongly on the model considered, and a dedicated fit would be required to make Eq.~\eqref{eq:vPhibound} quantitative.

\subsubsection*{The $\rho$ parameter}

In the limit $\delta \to 0$ the bidoublet VEV becomes proportional to the identity, $\langle \Phi \rangle \propto \id_2$, and the model possesses a residual global $SU(2)$ symmetry, discussed in App.~\ref{App:Symm}, which plays the role of a custodial symmetry. This symmetry is broken exclusively by the splitting $\delta$. Using the gauge boson masses of Eqs.~\eqref{eq:Zmasses} and \eqref{eq:Wmasses}, and noting that $M_Z^2$ receives no correction in $\delta$ while $M_W^2$ is shifted by $\frac{1}{2}g^2 v_\Phi^2 \, \delta^2$, we find the simple tree-level result
\begin{equation} \label{eq:rho}
\rho - 1 \simeq \frac{2 \, \delta^2}{\varepsilon^2} \, ,
\end{equation}
which involves only the two dimensionless expansion parameters and is independent of the gauge couplings. It is worth stressing that the gauge boson mixing does not contribute to $\rho$: the residual $SU(2)$ symmetry shifts $M_Z$ and $M_W$ by the same relative amount, and the effect cancels in the ratio.

Taken at face value, Eq.~\eqref{eq:rho} would constrain the ratio $\delta/\varepsilon$. However, $\delta$ is not a free parameter: it is fixed by minimizing the potential, Eq.~\eqref{eq:deltavalue}. Inserting that result,
\begin{equation}
\rho - 1 \simeq \frac{2 \, M_\eta^4 \, v_\eta^4}{\lambda_\Phi'^2 \, v_H^2 \, v_\Phi^6} \, ,
\end{equation}
which for the reference values $v_\Phi = 10$ TeV, $M_\eta \sim 1$ TeV, $v_\eta \sim 1$ GeV and $\lambda_\Phi' = 0.4$ gives $\rho - 1 \sim 2 \times 10^{-16}$, twelve orders of magnitude below the current experimental precision. Saturating the experimental bound would require $v_\eta \sim 10^3$ GeV, well above the electroweak scale and in sharp conflict with the hierarchy $v_\eta \ll v_H$ demanded by the neutrino mass mechanism. Therefore, the model automatically satisfies the constraints from the $\rho$ parameter, because the vacuum structure enforces an extremely well aligned bidoublet. The same conclusion applies to the condition $\lambda_\Phi' v_\Phi^2 \delta \ll \lambda_H v_H^2$ required for the validity of the scalar expansion, which is satisfied with a huge margin.

The only electroweak precision observable with real constraining power is therefore the gauge boson mixing discussed above, which bounds $v_\Phi$ from below.

\subsection{Collider signatures} \label{subsec:collider}

\subsubsection*{Nearly degenerate $W'$ and $Z'$ bosons}

As shown in Eqs.~\eqref{eq:Zmasses} and \eqref{eq:Wmasses}, the heavy gauge bosons satisfy $M_{W'}^2 = M_{Z'}^2 - \frac{1}{2} g^2 v_\Phi^2 \, \delta^2$, so that they are degenerate up to the small breaking induced by $\delta$. Their degeneracy is a sharp prediction of the model, again a consequence of the residual $SU(2)$ symmetry.

Since only the BSM leptonic states are charged under $SU(2)_2$, the direct couplings of $W'$ and $Z'$ to SM fermions are suppressed by $g_1/g_2$, and the standard Drell-Yan bounds are relaxed. Moreover, through the mixings of Eqs.~\eqref{eq:xiZ} and \eqref{eq:xiW}, the heavy gauge bosons inherit the couplings of their SM partners with a strength $\propto \xi_{Z,W}$, so that their leptophilic character is only approximate.

The prospects for a direct observation are, however, not straightforward, because the two suppression mechanisms work against each other. For $g_1 \sim g_2$ the direct coupling of the heavy gauge bosons to SM fermions is not suppressed at all and $Z'$ production proceeds essentially as for a sequential boson; but this is precisely the regime in which the gauge mixing of Eq.~\eqref{eq:xiZ} is largest, pushing $v_\Phi$ to the multi-TeV range and hence $M_{Z'} = \sqrt{(g_1^2+g_2^2)/2} \, v_\Phi$ into the region already probed by the dilepton searches at ATLAS and CMS. In the opposite regime, $g_2 \gg g_1$, the mixing is naturally suppressed and $v_\Phi$ may be lower, but the production cross section is then suppressed by $(g_1/g_2)^2$ and the heavy gauge bosons become very hard to produce. This tension has been studied in detail in Ref.~\cite{Cao:2012ng} for the general $SU(2)_1 \times SU(2)_2 \times U(1)_X$ framework, with the conclusion that the discovery reach at the LHC is severely limited by the constraints already imposed by low-energy precision data. A direct discovery in Drell-Yan therefore appears unlikely, and the role of these searches in the present model is to constrain the parameter space rather than to offer a discovery channel.

There is, on the other hand, a class of final states that is specific to our construction and to which the standard dilepton analyses are blind. Since the BSM leptonic states are the only fields charged under $SU(2)_2$, the decays $Z' \to \chi^+ \chi^-, \, \chi^0 \chi^0$ and $W' \to \chi^\pm \chi^0$ are unsuppressed and, for $M_{Z'} > 2 M_\chi$, dominate the total width. This has two consequences. First, it dilutes $\text{Br}(Z' \to \ell^+ \ell^-)$ and therefore relaxes the dilepton bounds with respect to the sequential case. Second, it leads to multilepton final states accompanied by missing energy and without hard jets. Finally, the near degeneracy $M_{W'} \simeq M_{Z'}$ discussed above, which is also found in Ref.~\cite{Cao:2012ng} for models following the second breaking pattern, would constitute a sharp test of the model should both states be observed.

\subsubsection*{Invisible Higgs decays}

Since the majoron is massless and escapes detection, the decay $h \to JJ$ contributes to the invisible Higgs width. We define the interaction Lagrangian between the Higgs boson and a pair of majorons as
\begin{equation}
\mathcal{L}_{hJJ} = \frac{1}{2}\,g_{hJJ}\,h\,J^2 \, ,
\end{equation}
where $g_{hJJ}$ is a dimensionful coupling whose value depends on the details of the model. The corresponding invisible decay width is
\begin{equation}
\Gamma(h \to JJ) = \frac{g_{hJJ}^2}{32 \,\pi \,m_h} \,.
\end{equation}
In our model, expanding the scalar potential around the vacuum and expressing the scalar fields in terms of the physical mass eigenstates, we obtain
\begin{equation} \label{eq:ghJJ}
g_{hJJ} \simeq \frac{\lambda_{H\Phi}\,m_h^2\,v_H}{\left(2\lambda_\Phi+\lambda_\Phi'\right)v_\Phi^2} \,.
\end{equation}
The coupling is suppressed by $m_h^2/v_\Phi^2$ with respect to the naive estimate $\lambda_{H\Phi} v_H$: the SM Higgs doublet carries no $B-L$ charge, and couples to the majoron only through its small mixing with $\Phi$, itself of order $v_H/v_\Phi$.

The resulting branching ratio is far below any foreseeable sensitivity: for $\lambda_{H\Phi} = 0.03$ one finds $\text{Br}(h \to JJ) \sim 10^{-6}$ at $v_\Phi = 3$ TeV and $\sim 10^{-8}$ at $v_\Phi = 10$ TeV. Even for $\lambda_{H\Phi} \sim 1$ and $v_\Phi$ at its lower limit the branching ratio stays around $10^{-3}$. This is a distinctive feature of the present construction. In singlet majoron models the scale $v_J$ at which lepton number is broken is a free parameter and may lie close to the electroweak scale, leading to sizable invisible Higgs decays~\cite{Joshipura:1992hp}. Here the majoron resides in the multiplet that breaks the gauge symmetry $SU(2)_1 \times SU(2)_2 \to SU(2)_L$, so that $v_\Phi$ is bounded from below by gauge boson mixing and by direct searches. Tying the majoron to the gauge symmetry breaking thus renders it invisible in Higgs decays.

\subsection{Lepton flavor violation and majoron phenomenology} \label{subsec:lfv}
To analyze the LFV phenomenology, it is useful to reconsider Eq.~\eqref{eq:Yukawadecom}. As discussed above, this decomposition makes the different sectors manifest, and one might therefore expect the model to exhibit the signatures associated with all these sectors simultaneously. However, as discussed at the end of Sec.~\ref{subsubsec:neutralfermions}, despite the presence of sectors traditionally associated with low-scale seesaw mechanisms, the interplay between the different contributions suppresses the associated phenomenology by the smallness of the light neutrino masses. This can be understood from an EFT perspective. Assuming $v_\Phi \gg \mu$, we can integrate out all fields with masses at the scale $v_\Phi$ or above. These include all massive BSM scalar fields and the pseudo-Dirac pair formed by $N$ and $\chi_s$. We are thus left with an EFT in which $U(1)_{B-L}$ is already broken and $SU(2)_1 \times SU(2)_2$ is broken down to $SU(2)_L$. The field content of this EFT consists of the SM fields, supplemented by those of the type-III seesaw and the massless majoron. The conventional LFV operators arise at dimension six and are therefore suppressed by two powers of $v_\Phi$. The interplay between the different neutrino mass mechanisms pushes $v_\Phi$ to very high scales or the Yukawas to very small values (see Eq.~\eqref{eq:vphimnubound}), rendering these contributions completely negligible.

In contrast to conventional scenarios, however, our EFT contains an additional light field, the majoron, which gives rise to new exotic LFV signatures. In particular, the majoron allows for a new dimension-five operator of the form
\begin{equation}
    \mathcal{O}_J = \frac{c_{ij}}{v_\Phi}\, \partial_\mu J \, \bar{L}_i \, \gamma^\mu L_j \, ,
\end{equation}
which induces new LFV processes that are not suppressed by two powers of $v_\Phi$. Indeed, using the equations of motion, one finds
\begin{equation}
    \mathcal{O}_J \sim J \, \bar{\ell}_i \, c_{ij} \frac{M_\ell}{v_\Phi}\, \ell_j \, .
\end{equation}
Thus, the majoron provides a distinctive avenue for potentially observable LFV in future experiments.

\subsubsection*{$\boldsymbol{\ell_\alpha \to \ell_\beta \, J}$}
As discussed above, this flavor-violating process arises from a dimension-five operator in the EFT and corresponds to a one-loop process in the full theory~\cite{Herrero-Brocal:2023czw}. However, a tree-level coupling between the majoron and charged leptons also exists, originating from the majoron components along the $H$ and $\eta$ directions. Nevertheless, since the EFT analysis shows that the one-loop coupling scales as $M_\ell/v_\Phi$, whereas the majoron components along the $H$ and $\eta$ directions are more strongly suppressed (see Eq.~\eqref{eq:Jprofile}), we expect the tree-level contribution to be subdominant. In particular, using the majoron profile in Eq.~\eqref{eq:Jprofile}, we find the following tree-level contribution:
\begin{align}
-\mathcal{L}_{J \mathcal{F}^\pm \mathcal{F}^\pm}^{\textup{tree-level}}\simeq \, \frac{i}{\sqrt{2}} \, \overline{\mathcal{F}}^\pm_L\,
\begin{pmatrix}
    \frac{v_\eta^2}{\sqrt{2} \, v_H v_\Phi} y_e & \frac{-v_\eta}{\sqrt{2} v_\Phi}\\
    0 & 0
\end{pmatrix}
\,\mathcal{F}^\pm_R+\hc\, ,
\end{align}
which, upon rotating to the charged-lepton mass basis, yields
\begin{equation}
-\mathcal{L}_{J \ell \ell}^{\textup{tree-level}} \simeq \frac{i}{\sqrt{2}} \, J \, \bar{\ell} \,  \left( \frac{v_\eta^2}{ v_H^2 v_\Phi} \,M_\ell + \frac{v_\eta^2}{2v_\Phi} \, M_\ell \, y_\chi \, \mu^{-2} \right) \, \ell   \, .
\end{equation}
Thus, the tree-level coupling is highly suppressed compared with the one-loop contribution.\\

To derive the analytical expression for the one-loop contribution~\cite{Herrero-Brocal:2023czw}, we first note that all contributions to the amplitude arising from the triplet sector are subdominant. This can be understood from the fact that the triplet sector decouples in the limit $v_\eta \to 0$ (which also implies $\delta \to 0$). Furthermore, since NGBs couple derivatively, the resulting coupling must be proportional to the charged-lepton mass matrix, with an overall scaling of $M_\ell/v_\Phi$. Consequently, contributions from the triplet sector must be further suppressed, relative to this leading scaling, by powers of $v_\eta/\Lambda$, with $\Lambda \gg v_\eta$, since $v_\eta$ corresponds to the lightest scale in the theory. It is therefore sufficient to consider the contribution from the singlet sector. This sector is equivalent to a theory based on the SM gauge group extended by a global $B-L$ symmetry, with two fermionic singlets, $N$ and $\chi_s$, a scalar singlet, $\phi_s$, and a scalar doublet, $\eta$. This model was studied by some of us in Ref.~\cite{CentellesChulia:2024uzv}, where it was denoted as model $C2b$. The corresponding coupling was found to be
\begin{align}
\label{eq:Jll1loop}
\mathcal{L}_{\ell\ell J} ={}& -\frac{iJ}{32\sqrt{2}\,\pi^2 v_\Phi}\,
\bar{\ell}\, \Big\{ M_\ell\,\textup{Tr}(y_N y_N^\dagger)\,\gamma_5
+ M_\ell(2y_N y_N^\dagger-y_\chi \Theta y_\chi^\dagger)P_L
\nonumber \\
&\qquad \qquad \qquad \qquad -(2y_N y_N^\dagger-y_\chi \Theta y_\chi^\dagger)M_\ell P_R
\Big\} \, \ell \, ,
\end{align}
with the matrix $\Theta$ defined by
\begin{equation}
    \Theta_{sp} \equiv \frac{(M_{N \chi_s}^2)_s}{\left( (M_{N \chi_s}^2)_s - M_{\eta}^2 \right)^2}  \left( (M_{N \chi_s}^2)_s- M_{\eta}^2 + M_{\eta}^2 \log \frac{M_{\eta}^2}{(M_{N \chi_s}^2)_s} \right) \delta_{sp} \, ,
\end{equation}
where $M_{N \chi_s} \equiv Y v_\Phi/\sqrt{2}$. Finally, let us note that, when comparing with the model of Ref.~\cite{CentellesChulia:2024uzv}, the $SU(2)$ structure leads to different conventions for the couplings. In particular, for the simplified model, the relevant terms in the Lagrangian are
\begin{align} 
-\mathcal{L}
= \; \; &\, y_N \, \overline{L} \, \tilde{H} \, N +y_\chi\, \overline{L}\,\tilde{\eta}\,\chi_s + Y \,\phi_s^\dagger \, \overline{N^c} \,\chi_s + \frac{\mu}{2} \,\overline{\tilde{\chi}}_s\,\chi_s  \, .
\end{align}
with $\phi_s = \frac{1}{\sqrt{2}} \left(v_{\phi_s} + S + i J \right)$. In our case, however, the $SU(2)$ structure implies the replacements $y_\chi \to y_\chi/\sqrt{2}$, which comes from the singlet projection of the
bidoublet $\chi$, and $v_{\phi_s} \to \sqrt{2}\, v_\Phi$, since $\langle \phi_s \rangle = v_\Phi$. Finally, the sign of $J$ is conventional and does not affect the rates.

We can now analyze the result. As expected from our EFT discussion, the majoron LFV phenomenology is dictated by the ratio $M_\ell/v_\Phi$, thus being enhanced with respect to the traditional LFV signals. Nevertheless, if we compare with the experimental limits, we find that even if this interaction is only suppressed by one power of $v_\Phi$, one still needs $v_\Phi$ not being very large in order to obtain sizable observables. This can be seen by taking into account the experimental constraints. Writing the $J \ell \ell$ interaction as~\cite{Escribano:2020wua}
\begin{equation} \label{eq:Jllgeneral}
\mathcal{L}_{\pi \ell \ell } = J \, \bar{\ell}_\beta \left( S_L^{\beta \alpha} \, P_L + S_R^{\beta \alpha} \, P_R \right) \ell_{\alpha} + \hc = J \, \bar{\ell}_\beta \left[ S^{\beta \alpha} \, P_L + \left( S^{\alpha \beta} \right)^* \, P_R \right] \ell_{\alpha} \, ,
\end{equation}
where all flavor combinations are considered, astrophysical observations impose strong constraints on the diagonal entry $S^{ee}$. In particular, energy-loss arguments lead to~\cite{DiLuzio:2020wdo,Bollig:2020xdr,Calibbi:2020jvd,Croon:2020lrf,DiLuzio:2020jjp,Caputo:2021rux}
\begin{equation} \label{eq:See}
\left| \text{Im} \, S^{e e} \right| < 2.1 \times 10^{-13} \, .
\end{equation}
On the other hand, off-diagonal entries are relevant for LFV processes. In particular, $S$ enters the $\ell_\alpha \to \ell_\beta \, J$ decay width as
\begin{equation} \label{eq:gammaNGB}
\Gamma(\ell_\alpha \to \ell_\beta \, J) = \frac{m_\alpha}{32 \, \pi} \, \left| \widetilde S^{\beta \alpha} \right|^2 \, ,
\end{equation}
with $\widetilde S$ given by
\begin{equation}
\left| \widetilde S^{\beta \alpha} \right| = \left( \left| S^{\beta \alpha}_L \right|^2 + \left| S^{\beta \alpha}_R \right|^2 \right)^{1/2} \, .
\end{equation}
Thus, the experimental limit~\cite{Hirsch:2009ee}
\begin{equation} \label{eq:BRmueJ}
\BR \left( \mu \to e \, J \right) \lesssim 10^{-5}
\end{equation}
translates into
\begin{equation} \label{eq:mulim}
\left| \widetilde S^{e \mu} \right| < 5.3 \times 10^{-11} \, .
\end{equation}
Future experiments such as Mu3e~\cite{Hesketh:2022wgw,Perrevoort:2024qtc} and COMET~\cite{COMET:2018auw,Xing:2022rob} will improve these constraints. In particular, the expected sensitivities for a massless NGB are
\begin{align}
    \BR\left(\mu \to e \, J \right)_\text{Mu3e} &\lesssim 6 \cdot 10^{-7} \quad \Rightarrow \quad \left| \widetilde S^{e \mu} \right| < 1.3 \times 10^{-11} \, ,
    \\[0.25cm]
    \BR\left(\mu \to e \, J \right)_\text{COMET} &\lesssim 4.6 \cdot 10^{-9} \quad \Rightarrow \quad \left| \widetilde S^{e \mu} \right| < 1.1 \times 10^{-12} \, . \label{eq:Comet}
\end{align}
Combining the experimental limits and future sensitivities with Eq.~\eqref{eq:Jll1loop}, and assuming Yukawa couplings of order $\mathcal{O}(1)$ to maximize the signal, we can estimate the range of $v_\Phi$ for which $\mu \to e\,J$ could be observable. The astrophysical constraints impose a lower bound,
\begin{align}
    v_\Phi \gtrsim 10^4 \; \textup{TeV} \, ,
\end{align}
whereas the projected sensitivity of COMET translates into an upper bound on $v_\Phi$ for which $\mu \to e\,J$ could be observed,
\begin{align}
    v_\Phi \lesssim 10^{5-6} \; \textup{TeV} \, .
\end{align}
However, a lower bound on $v_\Phi$ also follows from the requirement of reproducing the observed neutrino masses, as given in Eq.~\eqref{eq:vphimnubound}. Remarkably, this lower bound, which ensures that the model can consistently account for neutrino masses, coincides with the upper bound on $v_\Phi$ for which LFV signals remain observable. Furthermore, since the lower bound in Eq.~\eqref{eq:vphimnubound} is obtained by taking $\mu$ at the scale probed by current experimental constraints, this implies that observing $\mu \to e\,J$ would require the BSM charged leptons to lie close to the currently explored mass range, making them accessible to upcoming experiments. This provides a distinctive signature of our model, setting it apart from other scenarios that predict observable $\mu \to e\,J$ rates, such as the singlet model studied in Ref.~\cite{CentellesChulia:2024uzv}. Thus, even if the triplet does not directly participate in any observable process, its existence is crucial for testing and potentially ruling out the model. Indeed, since both neutrino mass constraints and majoron phenomenology push $v_\Phi$ to very high scales, the phenomenology of the additional gauge bosons, and thus of the structure related to the extra $SU(2)$ symmetry, becomes negligible. However, the interplay between neutrino masses, the masses of the BSM charged leptons, and the $\mu \to e\,J$ process leads to a particularly promising phenomenological scenario to falsify the underlying $SU(2)_1 \times SU(2)_2 \times U(1)_Y \times U(1)_{B-L}$ structure. 

Of course, the requirement that $\mu \sim 0.1$--$1$ TeV for $\mu \to e\,J$ to be observable can be relaxed for smaller Yukawa couplings. However, this typically renders the LFV process unobservable, except in a very restricted region of parameter space where $\mu \to e\,J$ remains within experimental reach while the BSM charged leptons are too heavy to be detected. This can be understood as follows. As $y_N$ decreases, its contributions to both neutrino masses and $\mu \to e\,J$ become smaller. However, the one-loop coupling also receives a contribution independent of $y_N$, proportional to $y_\chi\,\Theta\,y_\chi^\dagger$. For $M_{N\chi_s} \gtrsim M_\eta$, one has $\Theta \simeq 1$, and the LFV amplitude is therefore dominated by a contribution scaling as $y_\chi^2 M_\ell/v_\Phi$. As $y_N$ decreases, the lower bound on $v_\Phi$ derived from neutrino masses in Eq.~\eqref{eq:vphimnubound} is relaxed, potentially allowing larger values of $\mu$ while keeping $v_\Phi$ low enough for $\mu \to e\,J$ to remain observable. However, this possibility is severely constrained. Indeed, for $y_\chi = \mathcal{O}(1)$, the type-III seesaw contribution can become significant, and keeping it below the observed neutrino mass scale requires a strong suppression of $v_\eta^2/\mu$. Unless $v_\eta$ is extremely small, this pushes $\mu$ to very high scales and, consequently, raises $v_\Phi$ to values for which $\mu \to e\,J$ becomes unobservable.

We can estimate the extent to which this possibility is viable by combining the experimental sensitivity of COMET with the constraints from neutrino masses. For $\mu \to e\,J$ to be observable, the effective coupling must satisfy
\begin{align}
\label{eq:condCOMET}
    \tilde{S}_{e\mu}^{\textup{COMET}}
    \lesssim
    \frac{2 y_N^2-y_\chi^2\,\Theta}{32\sqrt{2}\pi^2}
    \frac{m_\mu}{v_\Phi}
    \quad \Rightarrow \quad
    v_\Phi \lesssim
    \frac{2 y_N^2-y_\chi^2\,\Theta}
    {32\sqrt{2}\pi^2\,\tilde{S}_{e\mu}^{\textup{COMET}}}
    m_\mu \, .
\end{align}
Combining this constraint with the one arising from the inverse seesaw contribution to neutrino masses, given in Eq.~\eqref{eq:vphiboundinverse}, we obtain
\begin{align}
\label{eq:condmu}
    \mu \lesssim \left( \frac{Y}{y_N} \frac{2 y_N^2-y_\chi^2\,\Theta} {32\sqrt{2}\pi^2\,\tilde{S}_{e\mu}^{\textup{COMET}}} \right)^2 \frac{m_\mu^2}{v_H^2}\,m_\nu \lesssim \frac{Y^2}{y_N^2} \left(2 y_N^2-y_\chi^2\,\Theta\right)^2  \,10^{-1}\,\textup{TeV} \, .
\end{align}
For Yukawa couplings of order unity, this constraint requires $\mu$ to lie near the mass range accessible to future experiments, or even within the range already excluded by current searches, in agreement with our previous conclusion. For $y_N \ll 1$, however, the bound allows for heavier charged leptons. This possibility is further constrained by requiring the type-III seesaw contribution to remain consistent with the observed neutrino masses. In particular, we require
\begin{equation}
    m_\nu \gtrsim y_\chi^2 \frac{v_\eta^2}{\mu} \, ,
\end{equation}
where the inequality expresses the requirement that the type-III contribution not exceed the observed neutrino mass scale. Combining this condition with the upper bound on $\mu$ in Eq.~\eqref{eq:condmu}, we obtain
\begin{align}
\label{eq:condveta}
    v_\eta  \lesssim \frac{Y}{y_\chi\,y_N} \frac{2 y_N^2-y_\chi^2\,\Theta} {32\sqrt{2}\pi^2\,\tilde{S}_{e\mu}^{\textup{COMET}}} \frac{m_\mu}{v_H}\,m_\nu \lesssim \frac{Y}{y_N} \left(2 y_N^2-y_\chi^2\,\Theta\right) \,10^{-1}\,\textup{MeV} \, .
\end{align}
Therefore, realizing an observable $\mu \to e\,J$ signal while keeping the BSM charged leptons beyond experimental reach requires not only a small $y_N$ and a sizable $y_\chi$, but also a very small $v_\eta$. This considerably restricts the viable parameter space, as it requires simultaneously small $y_N$ and $v_\eta$, together with sizable $y_\chi$ and $Y$. Furthermore, although such a hierarchy of Yukawa couplings is possible, it departs from the original motivation of seesaw mechanisms.

\begin{figure}[t]
    \centering

    \begin{subfigure}{0.7\textwidth}
        \centering
        \includegraphics[width=\linewidth]{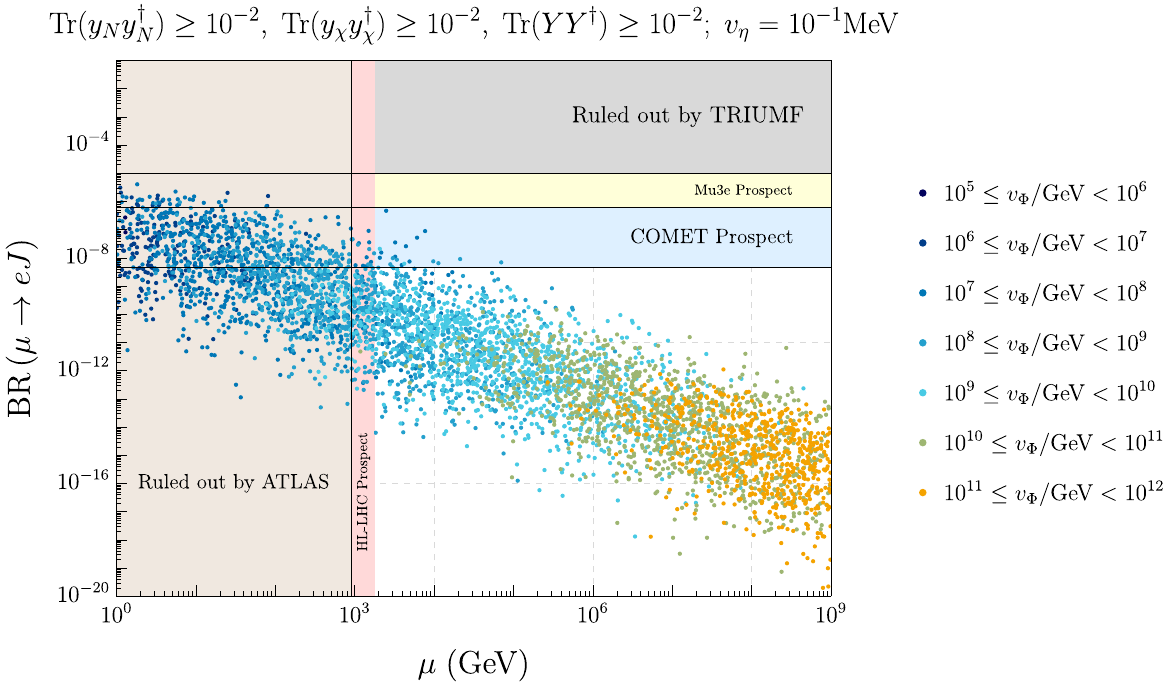}
        \caption{}
        \label{fig:plot1}
    \end{subfigure}
     \begin{subfigure}{0.48\textwidth}
        \centering
        \includegraphics[width=\linewidth]{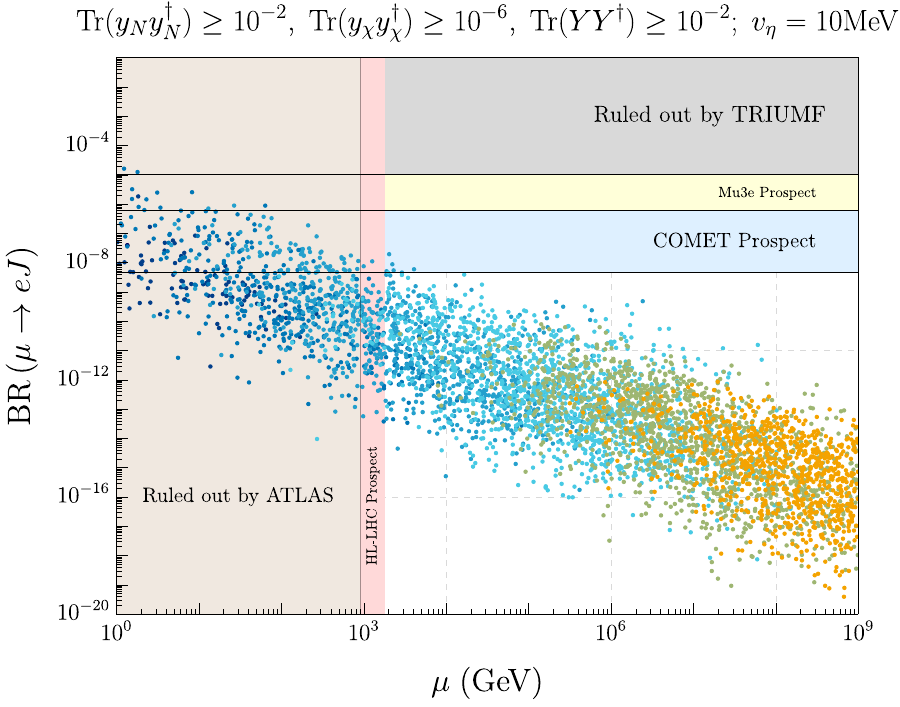}
        \caption{}
        \label{fig:plot2}
    \end{subfigure}
    \begin{subfigure}{0.48\textwidth}
        \centering
        \includegraphics[width=\linewidth]{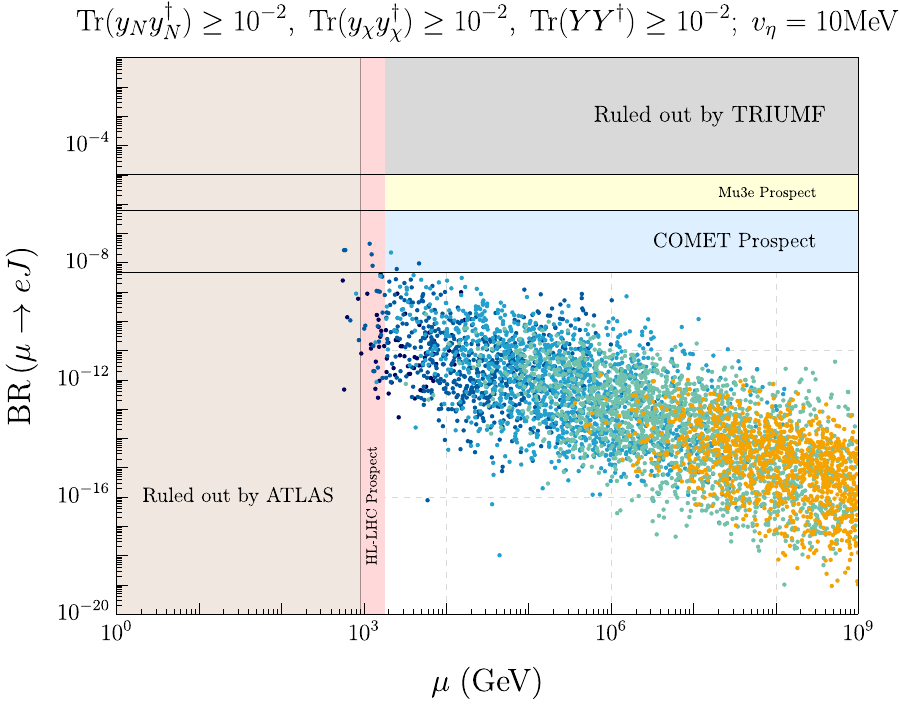}
        \caption{}
        \label{fig:plot3}
    \end{subfigure}
    \begin{subfigure}{0.48\textwidth}
        \centering
        \includegraphics[width=\linewidth]{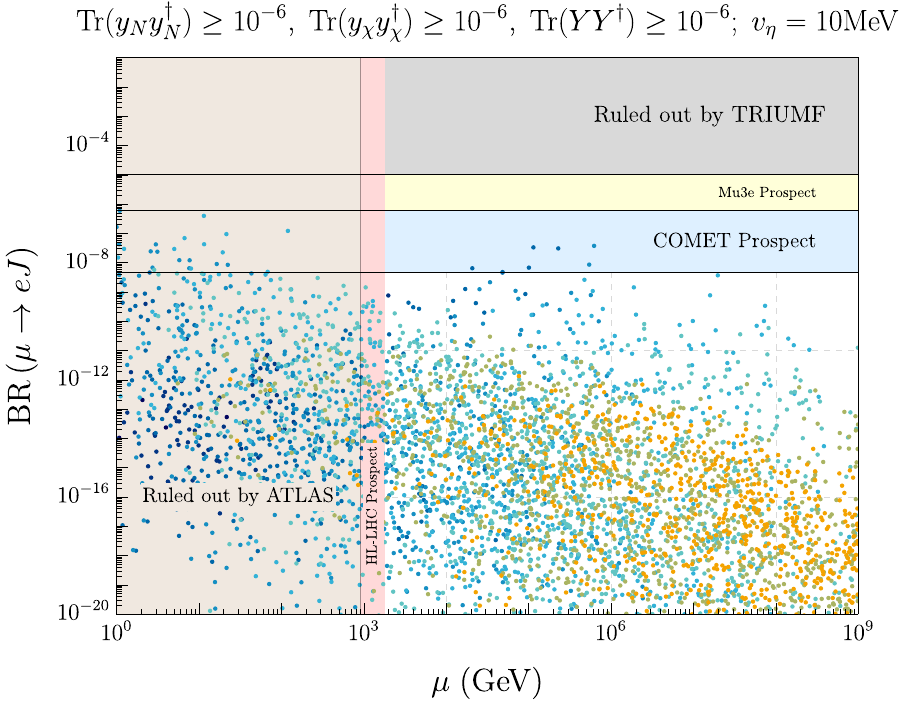}
        \caption{}
        \label{fig:plot4}
    \end{subfigure}
    \begin{subfigure}{0.48\textwidth}
        \centering
        \includegraphics[width=\linewidth]{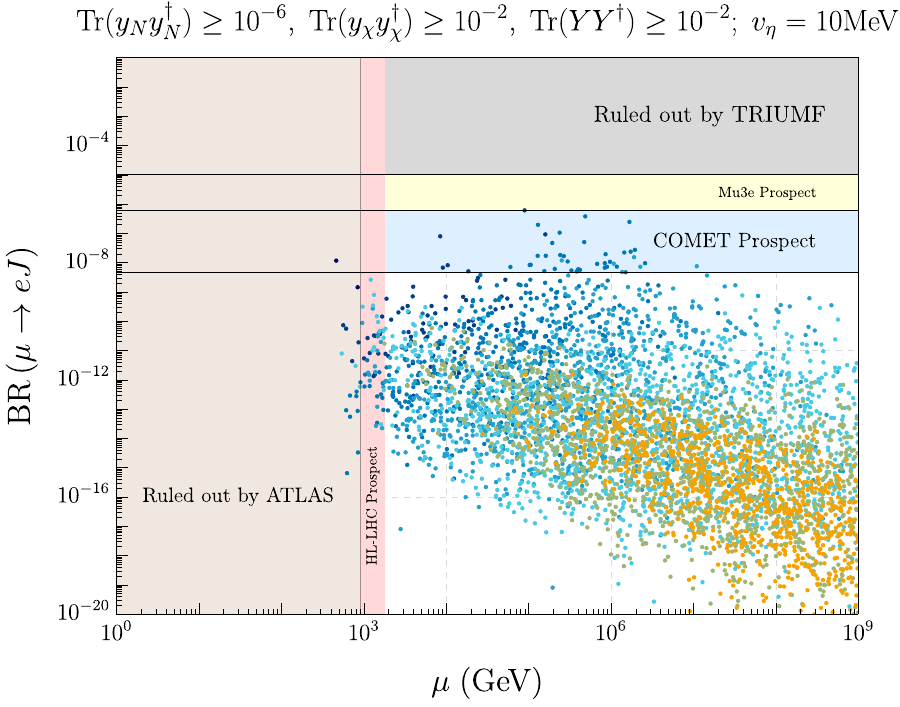}
        \caption{}
        \label{fig:plot5}
    \end{subfigure}
    \caption{$\BR(\mu \to e\,J)$ as a function of $\mu$ for different choices of the remaining parameters. The parameter regimes considered in each case are indicated in the title of the corresponding subfigure.}
    \label{fig:plots}
\end{figure}

These estimates can be further illustrated with a more general numerical analysis. In Fig.~\ref{fig:plots} we show the relation between $\BR(\mu \to e \, J)$ and the allowed scale for $\mu$. In the numerical scan used to generate these figures, all points reproduce the best-fit values of the neutrino oscillation parameters for normal ordering~\cite{deSalas:2020pgw} and comply with the conservative cosmological bound on the sum of neutrino masses, $\sum m_\nu < 0.16$~eV~\cite{DESI:2025zgx}. Furthermore, we require perturbative Yukawa couplings and impose the seesaw conditions and the astrophysical constraint of Eq.~\eqref{eq:See}. We fix the scale of $v_\eta$ while we vary the Yukawas $Y$, $y_N$ and $y_\chi$, as well as $v_\Phi$, in different ranges. For $M_\eta$ we take $M_\eta = 10^{-1}/\sqrt{2} v_\Phi$ in such a way that $\Theta \sim 1$. For the experimental constraints on the charged-fermion masses, we assume that the limits from conventional type-III seesaw searches apply. Thus, we take $\mu \simeq 0.9$ TeV for the current limit~\cite{ATLAS:2022yhd} and $\mu \simeq 1.8$ TeV as a optimistic prospect~\cite{Ruiz:2015zca}. However, as already mentioned, these limits should be regarded as estimates, since a dedicated analysis is required to establish their validity for this particular realization. 

As can be seen from the figures, our conclusions remain valid in the more detailed numerical analysis. Figures~\ref{fig:plot1} and~\ref{fig:plot2} show that, for sizable Yukawas satisfying Eq.~\eqref{eq:condveta}, observing $\mu \to e \, J$ implies a reachable mass scale for the BSM charged fermions. In contrast, Fig.~\ref{fig:plot3} shows that, even for large Yukawas, if Eq.~\eqref{eq:condveta} is not satisfied, $\mu \to e \, J$ is not observable, since neutrino masses push $\mu$, and therefore $v_\Phi$, to very high scales. Figure~\ref{fig:plot4} reveals that, when small Yukawas are allowed, the observation of $\mu \to e \, J$ still implies reachable charged leptons, except for the particular points where $y_N$ is very small while $y_\chi$ and $Y$ are sizable. Figure~\ref{fig:plot5} confirms that the points evading our general conclusion are precisely those with this particular hierarchy.
\\

\begin{figure}[t]
        \centering
        \includegraphics[width=0.7\textwidth]{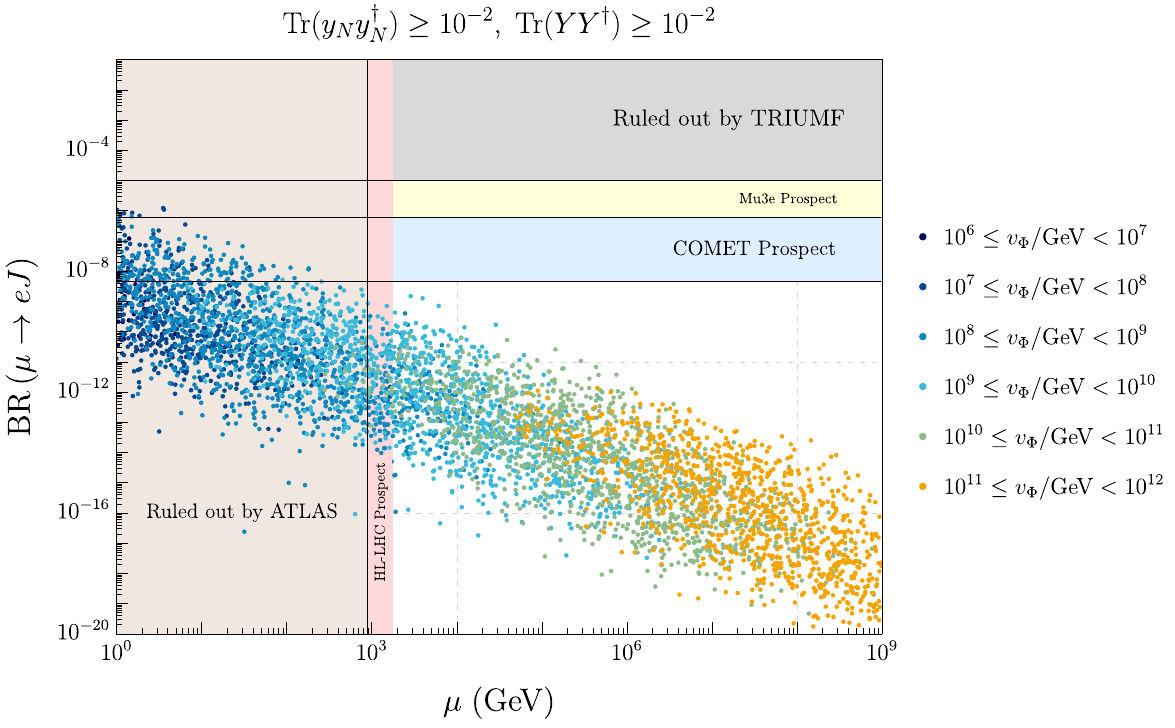}
        \caption{$\BR(\mu \to e\,J)$ as a function of $\mu$ for different choices of the remaining parameters in the limit where the doublet $\eta$ decouples.}
    \label{fig:plot6}
\end{figure}

Finally, let us comment on an interesting possibility. In our model we consider the most general hybrid scenario for the sake of generality. However, the $SU(2)_1 \times SU(2)_2 \times U(1)_{B-L}$ gauge structure offers a very interesting possibility in which it enforces a pure inverse seesaw. This corresponds to our model without the $SU(2)_2$ doublet $\eta$ or, equivalently, to the limit $y_\chi \to 0$. In this limit, the neutrino mass mechanism is a pure inverse seesaw, and the charged-scalar contribution to the majoron interaction with charged leptons disappears. In such a pure inverse seesaw scenario, we lose the freedom provided by $y_\chi$, and observing $\mu \to e \, J$ in future experiments would strongly constrain the model. Indeed, particularizing Eq.~\eqref{eq:condmu} for this case, we obtain
\begin{align}
    \mu  \lesssim Y^2 \, y_N^2  \; 10^{-1}\,\textup{TeV} \, .
\end{align}
Therefore, in the pure inverse seesaw scenario, an observation of $\mu \to e \, J$ would require the BSM charged leptons to lie at or below the TeV scale, within the reach of (or already in tension with) collider searches, making this realization of the model highly constrained. A more detailed numerical analysis can be found in Fig.~\ref{fig:plot6}.

Finally, being massless and coupled with strength $\sim 1/v_\Phi$, the majoron may contribute to $N_\text{eff}$. Requiring early enough decoupling yields a lower bound on $v_\Phi$, complementary to the one from gauge mixing. We refer to a similar analysis for the case of the diracon~\cite{CentellesChulia:2025eck} and leave this question to future work.
\section{Summary and conclusions}
\label{sec:Summ}

We have studied neutrino mass generation in a model based on the $\text{SU}(3)_c \times \text{SU}(2)_1 \times \text{SU}(2)_2 \times \text{U}(1)_Y$ gauge symmetry, in which the Standard Model $\text{SU}(2)_L$ emerges at low energies
as the diagonal subgroup of the two $\text{SU}(2)$ factors. The symmetry is extended with a global $\text{U}(1)_{B-L}$, spontaneously broken by the VEV of a scalar bidoublet $\Phi$, which also triggers the breaking $\text{SU}(2)_1 \times \text{SU}(2)_2 \to \text{SU}(2)_L$. The fermion sector is enlarged with singlets $N$ and real bidoublets $\chi$, and the scalar sector with an $\text{SU}(2)_2$ doublet $\eta$ whose small VEV, $v_\eta \ll v_H$, plays a key role in neutrino mass generation.

The light neutrino mass matrix receives three contributions, whose relative sizes are controlled by a single parameter, $r$, and stand in the ratio $1 : 2r : r^2$ for the type-III, linear and inverse terms, respectively. The type-III and inverse terms dominate for $r \ll 1$ and $r \gg 1$, respectively, whereas the linear term never dominates on its own. Despite the presence of matrix structures usually associated with low-scale seesaws, the parameter $\mu$ sets both the inverse seesaw scale and the mass of the type-III charged leptons. In practice, this implies that the light--heavy neutrino mixing is suppressed by the light neutrino mass, as in the canonical type-III seesaw, and reproducing the observed neutrino masses requires either very small Yukawa couplings or a high symmetry breaking scale, $v_\Phi \gtrsim (y_N/Y)\, 10^{5-6}\,\text{TeV}$ for
$\mu \gtrsim 0.1$--$1\,\text{TeV}$. The latter is also enforced by electroweak precision observables. In particular, by gauge boson mixing, which bounds the symmetry breaking scale from below, $v_\Phi \gtrsim 3\,\text{TeV}$ for $g_1 = g_2$.

Conventional lepton flavor violating processes arise at dimension 6 and are
therefore completely negligible for the values of $v_\Phi$ required by neutrino masses and gauge boson mixing. The majoron, however, induces a dimension-5 operator that gives rise to exotic $\ell_\alpha \to \ell_\beta J$ decays suppressed by a single power of $v_\Phi$, with couplings proportional to $M_\ell / v_\Phi$. These constitute the most promising signature of the model. Interestingly, we have shown that, for $\mathcal{O}(1)$ Yukawa couplings, an observation of $\mu \to e J$ would imply that the new charged leptons lie close to the mass range currently explored at colliders. This conclusion can only be evaded in a very limited region of the parameter space with very small $y_N$ and $v_\eta$ together with sizable $y_\chi$ and $Y$. The limit $y_\chi \to 0$, in which the model reduces to a pure inverse seesaw, is even more constrained. In all cases, the interplay between neutrino masses, the masses of the new charged leptons and $\mu \to e J$ provides a way to test, and potentially rule
out, the underlying $\text{SU}(2)_1 \times \text{SU}(2)_2 \times \text{U}(1)_Y \times \text{U}(1)_{B-L}$ structure.

Finally, several directions are left for future work. The massless majoron, coupled with strength $\sim 1/v_\Phi$, may contribute to $N_\text{eff}$ and lead to a lower bound on $v_\Phi$ complementary to the one from gauge boson mixing. A dedicated analysis of the collider limits on the type-III charged fermions, taking into account the more involved structure of this model, would also be required to turn our approximate bound on $\mu$ into a precise one.

\section*{Acknowledgements}

Work supported by the Spanish grants PID2023-147306NB-I00, CNS2024-154524 and CEX2023-001292-S (MICIU/AEI/10.13039/501100011033). J.P-S acknowledges support from the Spanish grant CIACIF/2022/158, funded by Generalitat Valenciana. A.H.B acknowledges support from the Spanish grant CIACIF/2021/100, funded by Generalitat Valenciana.

\appendix
\section{Recovering the SM $\boldsymbol{SU(2)_L}$} \label{App:Symm}

In this Appendix we show that the vacuum of the bidoublet $\Phi$ leaves an unbroken $SU(2)$ subgroup, which we identify with the SM $SU(2)_L$, and that this only happens for a specific alignment of its two VEVs. This justifies the symmetry breaking pattern assumed in Sec.~\ref{sec:Model}.

After a symmetry group $G$ is spontaneously broken, the subgroup that survives at low energies is the one that leaves the vacuum invariant,
\begin{equation}
    G' = \{\, g \in G \mid g \cdot \langle \Phi \rangle = \langle \Phi \rangle \,\} \, .
\end{equation}
In our case $G = SU(2)_1 \times SU(2)_2$ and the bidoublet transforms as $\Phi \to U_1 \, \Phi \, U_2^\dagger$, with $U_i \in SU(2)_i$. The invariance condition therefore reads
\begin{equation} \label{eq:invariance}
 U_1 \, \langle \Phi \rangle \, U_2^\dagger = \langle \Phi \rangle \; \Rightarrow \; U_1 \, \langle \Phi \rangle = \langle \Phi \rangle \, U_2 \, .
\end{equation}
This is a condition relating the two transformations: a given $U_1$ belongs to the unbroken subgroup only if a $U_2$ exists such that Eq.~\eqref{eq:invariance} holds. Our goal is then to determine for which vacua such a pairing is possible for every $U_1$, since only in that case a full $SU(2)$ survives.

Without loss of generality we can take $\langle \Phi \rangle$ to be diagonal, since any bidoublet VEV can be brought to this form by a suitable $SU(2)_1 \times SU(2)_2$ transformation. In order to eliminate $U_2$ from Eq.~\eqref{eq:invariance}, we multiply it by its Hermitian conjugate and use the unitarity of $U_2$,
\begin{equation}
 U_1 \, \langle \Phi \rangle \langle \Phi \rangle^\dagger \, U_1^\dagger = \langle \Phi \rangle \langle \Phi \rangle^\dagger \; \Rightarrow \; \left[ U_1 \, , \, \langle \Phi \rangle \langle \Phi \rangle^\dagger \right] = 0 \, .
\end{equation}
The matrix $\langle \Phi \rangle \langle \Phi \rangle^\dagger$ is diagonal, with entries $\left| \langle \Phi \rangle_{ii} \right|^2$. Writing the commutator in components,
\begin{equation}
 \left( U_1 \right)_{ij} \left[ \left| \langle \Phi \rangle_{ii} \right|^2 - \left| \langle \Phi \rangle_{jj} \right|^2 \right] = 0 \, ,
\end{equation}
and two cases arise:
\begin{itemize}

\item If $\left| \langle \Phi \rangle_{11} \right| \neq \left| \langle \Phi \rangle_{22} \right|$, the bracket does not vanish for $i \neq j$ and the off-diagonal entries of $U_1$ are forced to be zero. Only diagonal transformations survive, and the same reasoning applies to $U_2$. The unbroken subgroup is then not an $SU(2)$, but merely a $U(1)$, and the SM $SU(2)_L$ is not recovered.

\item If $\left| \langle \Phi \rangle_{11} \right| = \left| \langle \Phi \rangle_{22} \right|$, the bracket vanishes identically and the condition becomes empty: any $U_1$ is allowed, provided that $U_2$ is chosen accordingly. Solving Eq.~\eqref{eq:invariance} for $U_2$, the unbroken subgroup is
\begin{equation}
    G' = \{\, \left( U_1 \, , \, U_2 \right) \, \mid \; U_1 \in SU(2)_1 \, , \,  U_2 =\langle \Phi \rangle^{-1} \, U_1 \, \langle \Phi \rangle  \,\} \, ,
\end{equation}
which is isomorphic to $SU(2)$ and can be identified with the $SU(2)_L$ of the SM.

\end{itemize}
An exact $SU(2)_L$ is therefore obtained only for $\left| v_w \right| = \left| v_z \right|$, that is, for $\delta = 0$. Taking all the VEVs to be real and positive, this implies $\langle \Phi \rangle \propto \id_2$ and hence $U_2 = U_1$, so that the surviving group is precisely the diagonal subgroup, $SU(2)_\text{diagonal} \equiv SU(2)_L$. As discussed in Sec.~\ref{sec:Model}, the exact alignment $\delta = 0$ is not compatible with $\kappa \neq 0$. The minimisation of the scalar potential generates a small splitting, and $SU(2)_L$ is only an approximate symmetry of the model. This is the origin of the residual $SU(2)$ symmetry invoked throughout the paper.

\section{Seesaw Expansion}\label{App:Seesaw}
In this Appendix, we derive the conditions for the validity of the seesaw expansion and present the corresponding results. We begin by defining our general framework. We consider a general complex matrix of the form
\begin{equation}
\mathcal{M} =
\begin{pmatrix}
M_1 & M_2\\
M_3 & M_4
\end{pmatrix}
\end{equation}
where $M_1$, $M_2$, $M_3$, and $M_4$ are general $n \times n$, $n \times (n+m)$, $(n+m) \times n$, and $m \times m$ matrices, respectively. The matrix can always be diagonalized by means of two unitary matrices according to
\begin{equation}
U_L^\dagger \, \mathcal{M} \, U_R = \hat{M} \equiv \textup{diag}\left(M_1 , \dots , M_{n+m} \right) , .
\end{equation}
The unitary matrices can always be parametrized as
\begin{equation} \label{eq:Ufactor}
U_{L,R}=
\begin{pmatrix}
\sqrt{\id_n -  B_{L,R}B_{L,R}^\dagger} & B_{L,R}\\
-B_{L,R}^\dagger & \sqrt{\id_m-  B_{L,R}^\dagger B_{L,R}}
\end{pmatrix}
\begin{pmatrix}
V_{L,R} & 0\\
0& W_{L,R}
\end{pmatrix} \equiv (U_1)_{L,R} , (U_2)_{L,R}, .
\end{equation}
Here, $V_{L,R}$ and $W_{L,R}$ are $n \times n$ and $m \times m$ unitary matrices, respectively, and the square roots are understood as power expansions. With this definition, $(U_1)_{L,R}$ bring $\mathcal{M}$ into block-diagonal form, while $(U_2)_{L,R}$ diagonalize each block independently.

We can find the mixing matrices $B_{L,R}$ by imposing the block-diagonalization. The off-diagonal entries of the block-diagonalized matrix $(U_1)_L^\dagger \, \mathcal{M} \, (U_1)_R$ are given by
\begin{align}
    \sqrt{\id_n -  B_{L}B_{L}^\dagger} \left(M_1 \, B_R + M_2 \, \sqrt{\id_m -  B_{R}B_{R}^\dagger} \right) - B_L \left( M_3 \, B_R + M_4 \sqrt{\id_m -  B_{R}^\dagger B_{R}} \right) &= 0 \, ,
    \\
    B_L^\dagger \left( M_1 \sqrt{\id_n -  B_{L}B_{L}^\dagger} - M_2 \, B_R^\dagger \right) + \sqrt{\id_n -  B_{L}^\dagger B_{L}} \left( M_3 \,\sqrt{\id_n -  B_{R} B_{R}^\dagger } - M_4 B_R^\dagger \right) &= 0 \, .
\end{align}
These equations can be solved by assuming $\left(B_{L,R}\right)_{ij} \ll 1$, corresponding to small mixing between the two block-diagonalized sectors. This is precisely the seesaw limit. In this limit, the above equations become
\begin{align} 
    M_1 \, B_R + M_2 - B_L \left( M_4 +M_3 B_R \right) &= 0 \, , \label{eq:seesaw1}
    \\
    B_L^\dagger \, M_1 + M_3 - \left(M_4 -B_L^\dagger M_2\right) B_R^\dagger &= 0 \, .\label{eq:seesaw2}
\end{align}
Introducing these equations, into the diagonal blocks one finds the following block-diagonal form at leading order:
\begin{equation}
    U_L^\dagger \, \mathcal{M} \, U_R = (U_2)_{L}^\dagger (U_1)_{L}^\dagger \, \mathcal{M} \, (U_1)_{R}\,  (U_2)_{R} \simeq (U_2)_{L}^\dagger
    \begin{pmatrix}
        m-B_L\,M_3 & 0 \\
        0 & M_4 + M_3 \,B_R
    \end{pmatrix} (U_2)_{R}
\end{equation}
These equations define the seesaw expansion and provide the corresponding formulas for block diagonalization. However, for a given model, one must verify that the solutions for $B_{L,R}$ obtained from Eqs.~\eqref{eq:seesaw1} and~\eqref{eq:seesaw2} satisfy the seesaw conditions $\left(B_{L,R}\right)_{ij} \ll 1$. In principle, these equations can be solved component by component. In the applications considered in this work, however, as is typically the case in the literature, we focus on scenarios in which the equations admit further simplifications. We discuss these cases below.
\subsection{Scenario 1: $M_1 \ll M_4$}
In this case, one finds
\begin{align}
B_R &= M_3^\dagger \,\left(M_4^\dagger\right)^{-1}  \, ,\label{eq:BR}
\\
B_L &= M_1 \,M_4^{-1} \label{eq:BL} \,.
\end{align}
Hence, the seesaw expansion is consistent if and only if
\begin{align}
    \left( B_R \right)_{ij} &= \left[\left(M_3^\dagger \, M_4^\dagger\right)^{-1} \right]_{ij} \ll 1 \, ,
    \\
   \left( B_L\right)_{ij} &= \left( M_2 \,M_4^{-1} \right)_{ij} \ll 1 .
\end{align}
If these conditions are satisfied, the diagonalization is given by
\begin{equation}
    U_L^\dagger \, \mathcal{M} \, U_R  \simeq (U_2)_{L}^\dagger
    \begin{pmatrix}
        M_1 -M_2 \, M_4^{-1} \, M_3 & 0 \\
        0 & M_4
    \end{pmatrix} (U_2)_{R}
\end{equation}
This is the seesaw expansion used in Section~\ref{subsubsec:chargedfermions}. If $\mathcal{M}$ is symmetric, then $U_L^\dagger = U_R^T$ and $B_L^\dagger = B_R^T$. This corresponds to the standard seesaw framework for neutral fermions used in Section~\ref{subsubsec:neutralfermions}.
\subsection{Scenario 2: $\mathcal{M}\in \mathbb{R}^{(n+m)\times(n+m)}$ and $\mathcal{M}= \mathcal{M}^T$}
If the mass matrix is real and symmetric, then $U_L^\dagger = U_R^T$ and $B_R^\dagger = B_R^T \equiv B^T$. Equation~\eqref{eq:seesaw1} (which, in the symmetric case, is equivalent to Eq.~\eqref{eq:seesaw2}, since $M_1$ and $M_4$ are symmetric matrices and $M_2=M_3^T$) can then be written as
\begin{align} 
    M_1 \, B + M_2 - B \, M_4 \simeq 0 \, .
\end{align}
One can then take $M_1 \in \mathbb{R}$ to be a number rather than a matrix and perform the diagonalization block by block. In this case, $M_1$ commutes with $B$, and one obtains
\begin{align} 
     B  \simeq M_2 \, \left(M_1 \id_m - M_4 \right)^{-1} \, .
\end{align}
Thus, the seesaw expansion is consistent if and only if
\begin{align} 
     B_{ij}  \simeq \left[ M_2 \left(M_1 \id_m - M_4 \right)^{-1} \right]_{ij}\ll 1 \, .
\end{align}
In this case, the diagonalization is given by
\begin{equation}
    U_L^\dagger \, \mathcal{M} \, U_R  \simeq (U_2)_{L}^\dagger
    \begin{pmatrix}
        M_1 -M_2 \, \left(M_1 \id_m - M_4 \right)^{-1} \, M_2^T & 0 \\
        0 & M_4 + M_2^T \, M_2 \left(M_1 \id_m - M_4 \right)^{-1}\, 
    \end{pmatrix} (U_2)_{R}
\end{equation}
Provided that the seesaw condition, i.e.,
$\left[ M_2 \left(M_1 \id_m - M_4 \right)^{-1} \right]_{ij}\ll 1$,
is satisfied, the full matrix can be diagonalized by applying successive seesaw diagonalizations. This is precisely the procedure followed when diagonalizing the CP-even mass matrix.

\section{Scalar mass eigenstates}\label{App:Scalar}
\subsection{CP-even mass matrix} \label{subsec:CPeven}
In the original basis, $\mathcal{S}^T\equiv \left( S_H \ \ S_w \ \ S_z \ \ S_\eta \right)$, the mass matrix is given by
\begin{align}
\mathcal{M}_\mathcal{S}^2 &=
\begin{pmatrix}
\frac{2 v_H^{4} \lambda_H + (v_z^{2}- v_w^{2}) v_z^{2}\lambda_\Phi' }{2 v_H^{2}}
&
v_H v_w \lambda_{H\Phi}
&
v_H v_z \lambda_{H\Phi} + \frac{v_z (v_w^{2} - v_z^{2}) \lambda_\Phi'}{2 v_H}
&
\lambda_{H \eta} v_H v_\eta + \frac{v_z^{2} (v_w^{2} - v_z^{2}) \lambda_\Phi'}{2 v_H v_\eta}
\\
v_H v_w \lambda_{H\Phi}
&
v_w^{2} (\lambda_\Phi + \lambda_\Phi')
&
v_w v_z \lambda_\Phi
&
v_\eta v_w \lambda_{\eta\Phi}
\\
v_H v_z \lambda_{H\Phi} + \frac{v_z (v_w^{2} - v_z^{2}) \lambda_\Phi'}{2 v_H}
&
v_w v_z \lambda_\Phi
&
-\frac{v_w^{2} \lambda_\Phi'}{2} + v_z^{2} \left( \lambda_\Phi + \frac{3 \lambda_\Phi'}{2} \right)
&
v_\eta v_z \lambda_{\eta\Phi} + \frac{v_z (v_w^{2} - v_z^{2}) \lambda_\Phi'}{2 v_\eta}
\\
\lambda_{H \eta} v_H v_\eta + \frac{v_z^{2} (v_w^{2} - v_z^{2}) \lambda_\Phi'}{2 v_H v_\eta}
&
v_\eta v_w \lambda_{\eta\Phi}
&
v_\eta v_z \lambda_{\eta\Phi} + \frac{v_z (v_w^{2} - v_z^{2}) \lambda_\Phi'}{2 v_\eta}
&
\frac{2 v_\eta^{4} \lambda_\eta +(v_z^{2}- v_w^{2}) v_z^{2} \lambda_\Phi'}{2 v_\eta^{2}} 
\end{pmatrix} \; .
\end{align}
It is however more convenient to work in the basis where the fields transform irreducibly under $SU(2)_L$: $\mathcal{S}^T \equiv \left( S_H \ \ S_s \ \ S_t \ \ S_\eta \right)$. Since the full mass matrix cannot be diagonalized analytically, we expand it in terms of two small parameters, $\varepsilon \equiv  \,v_H/v_\phi$ and $\chi \equiv  \, v_\eta /v_\phi$. Given that $\varepsilon \gg \chi$, we truncate the expansion at $\mathcal{O}(\varepsilon^2)$ and $\mathcal{O}(\chi)$. To consistently handle terms such as $\chi^2/\varepsilon$, we introduce the auxiliary parameter $\xi$ as $\varepsilon \to \xi \varepsilon$, $\chi \to \xi \chi$ and expand up to second order in $\xi$. Consequently, the mass matrix takes the following form
\begin{align}
\mathcal{M}_\mathcal{S}^2 &=
\begin{pmatrix}
\frac{M_{\eta}^2 v_{\eta}^2}{v_H^2} + v_H^2 \lambda_H & \sqrt{2} v_H v_{\phi} \lambda_{H\phi} & 0 & -\frac{M_{\eta}^2 v_{\eta}}{v_H} \\[1.5ex]
\sqrt{2} v_H v_{\phi} \lambda_{H\phi} & v_{\phi}^2 \left(2 \lambda_{\phi} + \lambda_{\phi}'\right) & 0 & \frac{v_{\eta} \left(-M_{\eta}^2 + 2 v_{\phi}^2 \lambda_{\eta\phi}\right)}{\sqrt{2} v_{\phi}} \\[1.5ex]
0 & 0 & v_{\phi}^2 \lambda_{\phi}' & \frac{M_{\eta}^2 v_{\eta}}{\sqrt{2} v_{\phi}} \\[1.5ex]
-\frac{M_{\eta}^2 v_{\eta}}{v_H} & \frac{v_{\eta} \left(-M_{\eta}^2 + 2 v_{\phi}^2 \lambda_{\eta\phi}\right)}{\sqrt{2} v_{\phi}} & \frac{M_{\eta}^2 v_{\eta}}{\sqrt{2} v_{\phi}} & M_{\eta}^2
\end{pmatrix} \; .
\end{align}
Since the seesaw condition is satisfied for all submatrices (see App.~\ref{App:Seesaw}), we can perform the seesaw diagonalization block by block. Thus, applying successive seesaw diagonalizations, we obtain,
\begin{align}
&\left(\begin{smallmatrix}
\frac{M_{\eta}^2 v_{\eta}^2}{v_H^2} + v_H^2 \lambda_H & \sqrt{2} v_H v_{\phi} \lambda_{H\phi} & 0 & -\frac{M_{\eta}^2 v_{\eta}}{v_H} \\
\sqrt{2} v_H v_{\phi} \lambda_{H\phi} & v_{\phi}^2 \left(2 \lambda_{\phi} + \lambda_{\phi}'\right) & 0 & \frac{v_{\eta} \left(-M_{\eta}^2 + 2 v_{\phi}^2 \lambda_{\eta\phi}\right)}{\sqrt{2} v_{\phi}} \\
0 & 0 & v_{\phi}^2 \lambda_{\phi}' & \frac{M_{\eta}^2 v_{\eta}}{\sqrt{2} v_{\phi}} \\
-\frac{M_{\eta}^2 v_{\eta}}{v_H} & \frac{v_{\eta} \left(-M_{\eta}^2 + 2 v_{\phi}^2 \lambda_{\eta\phi}\right)}{\sqrt{2} v_{\phi}} & \frac{M_{\eta}^2 v_{\eta}}{\sqrt{2} v_{\phi}} & M_{\eta}^2
\end{smallmatrix}\right)
\Rightarrow
\left(\begin{smallmatrix}
\frac{M_{\eta}^2 v_{\eta}^2 + v_H^4 \lambda_H}{v_H^2} & \sqrt{2} v_H v_{\phi} \lambda_{H\phi} & 0 & -\frac{M_{\eta}^2 v_{\eta}}{v_H} \\[1.5ex]
\sqrt{2} v_H v_{\phi} \lambda_{H\phi} & v_{\phi}^2 \left(2 \lambda_{\phi} + \lambda_{\phi}'\right) & 0 & 0 \\[1.5ex]
0 & 0 & v_{\phi}^2 \lambda_{\phi}' & \frac{M_{\eta}^2 v_{\eta}}{\sqrt{2} v_{\phi}} \\[1.5ex]
-\frac{M_{\eta}^2 v_{\eta}}{v_H} & 0 & \frac{M_{\eta}^2 v_{\eta}}{\sqrt{2} v_{\phi}} & M_{\eta}^2
\end{smallmatrix}\right)
\nonumber \\[1cm]
\Rightarrow
&\left(\begin{smallmatrix}
\frac{M_{\eta}^2 v_{\eta}^2 + v_H^4 \lambda_H}{v_H^2} & \sqrt{2} v_H v_{\phi} \lambda_{H\phi} & 0 & -\frac{M_{\eta}^2 v_{\eta}}{v_H} \\[1.5ex]
\sqrt{2} v_H v_{\phi} \lambda_{H\phi} & v_{\phi}^2 \left(2 \lambda_{\phi} + \lambda_{\phi}'\right) & 0 & 0 \\[1.5ex]
0 & 0 & v_{\phi}^2 \lambda_{\phi}' & 0 \\[1.5ex]
-\frac{M_{\eta}^2 v_{\eta}}{v_H} & 0 & 0 & M_{\eta}^2
\end{smallmatrix}\right)
\Rightarrow
\left(\begin{smallmatrix}
v_H^2 \lambda_H & \sqrt{2} v_H v_{\phi} \lambda_{H\phi} & 0 & 0 \\[1.5ex]
\sqrt{2} v_H v_{\phi} \lambda_{H\phi} & v_{\phi}^2 \left(2 \lambda_{\phi} + \lambda_{\phi}'\right) & 0 & 0 \\[1.5ex]
0 & 0 & v_{\phi}^2 \lambda_{\phi}' & 0 \\[1.5ex]
0 & 0 & 0 & M_{\eta}^2 
\end{smallmatrix}\right) 
\nonumber 
\\[1cm]
\Rightarrow
&\left(\begin{smallmatrix}
\frac{v_H^2 \left( \lambda_H (2\lambda_{\phi} + \lambda_{\phi}') - 2\lambda_{H\phi}^2 \right)}{2 \lambda_{\phi} + \lambda_{\phi}'} & 0 & 0 & 0 \\[1.5ex]
0 & v_{\phi}^2 \left(2 \lambda_{\phi} + \lambda_{\phi}'\right) & 0 & 0 \\[1.5ex]
0 & 0 & v_{\phi}^2 \lambda_{\phi}' & 0 \\[1.5ex]
0 & 0 & 0 & M_{\eta}^2 
\end{smallmatrix}\right) \; .
\end{align} 
Consequently, the masses are determined by the mass matrix given above, while the mass eigenstates are obtained from the unitary matrix resulting from the full seesaw expansion. Using the definition of this unitary matrix in Eq.~\eqref{eq:Ufactor}, we find
\begin{align}
    \begin{pmatrix}
      \S_1 \\
      \S_2 \\
      \S_3 \\
      \S_4
    \end{pmatrix} & \simeq 
\left(\begin{smallmatrix}
1  & \frac{\sqrt{2} v_H \lambda_{H\phi}}{2 v_{\phi} \lambda_{\phi} + v_{\phi} \lambda_{\phi p}} & 0 & -\frac{v_{\eta}}{v_H} \\[1.5ex]
-\frac{\sqrt{2} v_H \lambda_{H\phi}}{2 v_{\phi} \lambda_{\phi} + v_{\phi} \lambda_{\phi p}} & 1  & 0 & \frac{v_{\eta} \left(M_{\eta}^2 + 2 v_{\phi}^2 (\lambda_{H\phi} - \lambda_{\eta\phi})\right)}{\sqrt{2} \left(-M_{\eta}^2 v_{\phi} + v_{\phi}^3 (2 \lambda_{\phi} + \lambda_{\phi p})\right)} \\[1.5ex]
0 & 0 & 1 & \frac{M_{\eta}^2 v_{\eta}}{\sqrt{2} \left(M_{\eta}^2 v_{\phi} - v_{\phi}^3 \lambda_{\phi p}\right)} \\[1.5ex]
\frac{v_{\eta}}{v_H} & \frac{2 v_{\eta} v_{\phi}^2 \lambda_{\eta\phi} (2 \lambda_{\phi} + \lambda_{\phi p}) - M_{\eta}^2 v_{\eta} (2 \lambda_{H\phi} + 2 \lambda_{\phi} + \lambda_{\phi p})}{\sqrt{2} v_{\phi} (2 \lambda_{\phi} + \lambda_{\phi p}) \left(-M_{\eta}^2 + v_{\phi}^2 (2 \lambda_{\phi} + \lambda_{\phi p})\right)} & \frac{M_{\eta}^2 v_{\eta}}{\sqrt{2} \left(-M_{\eta}^2 v_{\phi} + v_{\phi}^3 \lambda_{\phi p}\right)} & 1 
\end{smallmatrix}\right)
\begin{pmatrix}
      S_H \\
      S_w \\
      S_z \\
      S_\eta
    \end{pmatrix} \,  . 
\end{align}
\subsection{CP-odd mass matrix} \label{subsec:CPodd}

In the basis $\mathcal{A}^T \equiv \left( A_H \ \ A_w \ \ A_z \ \ A_\eta \right)$, the CP-odd mass matrix reads
\begin{equation} \label{eq:MCPodd}
\M_\P^2 = \frac{\lambda_\Phi' \left( v_z^2 - v_w^2 \right)}{2}
\begin{pmatrix}
\frac{v_z^{2}}{v_H^{2}} & 0 & -\frac{v_z}{v_H} & -\frac{v_z^{2}}{v_H v_\eta} \\
0 & 0 & 0 & 0 \\
-\frac{v_z}{v_H} & 0 & 1 & \frac{v_z}{v_\eta} \\
-\frac{v_z^{2}}{v_H v_\eta} & 0 & \frac{v_z}{v_\eta} & \frac{v_z^{2}}{v_\eta^{2}}
\end{pmatrix} \, .
\end{equation}
This matrix has rank one, implying that only one CP-odd state acquires a non-vanishing mass. In the basis $\mathcal{A}$, the corresponding massive eigenstate is
\begin{align}
    \P_1 &\simeq \mathcal{N}_{\P_1} \left(- \frac{v_\eta}{v_H} \ \ 0 \ \ \frac{v_\eta}{v_\Phi} \ \ 1 \right) \; , \qquad \textup{with mass} &&  M_{\P_1}^2 = \lambda_\Phi' \frac{v_\Phi^4}{v_\eta^2} \delta \, ,
\end{align}
where $\mathcal{N}_{\P_1}$ is a normalization factor. The remaining three orthogonal states are massless and correspond to NGBs of the theory. Two of them are absorbed by the $Z$ and $Z'$ gauge bosons, while the third remains in the physical spectrum as the majoron, $J$, associated with the spontaneous breaking of $U(1)_{B-L}$.

To identify the majoron, it is useful to derive the general conditions satisfied by a physical NGB: it must be massless and it must be orthogonal to the NGBs absorbed by the gauge bosons. To derive these conditions, let us consider a generic scalar potential of the form
\begin{align}
V(\phi_i) = \sum_\alpha V_\alpha(\phi_i) + \hc = \sum_\alpha \lambda_\alpha \, \phi_1^{n_{1,\alpha}}\,  \phi_2^{n_{2,\alpha}}  \dots \phi_N^{n_{N,\alpha}} + \hc \, ,
\end{align}
where $n_{i,\alpha}\in\mathbb{Z}$ denotes the power of the field $\phi_i$ in the interaction term $V_\alpha$. Here, $\phi_i$ denotes a generic scalar multiplet. It is convenient to use the polar parametrization of the scalar fields,
\begin{equation}\label{eq:polarparam}
\phi_j(x) = \frac{1}{\sqrt{2}}e^{i \frac{\theta_j(x)}{v_j}} \left( v_j+ S_j(x)\right)
= \frac{1}{\sqrt{2}} \left(v_j+ S_j(x) + i \theta_j(x) \right)
+ \mathcal{O}\left(\frac{\theta_j(x)}{v_j}\right) \, .
\end{equation}
The physical NGB is a linear combination of the phases (in the polar parametrization) or the CP-odd fields (in the linear parametrization), i.e.,
\begin{equation}
\pi(x)=\sum_j \beta_j\,\theta_j(x)\, .
\end{equation}
The condition that $\pi$ is massless can then be written as
\begin{equation} \label{eq:masslessNGB}
0= \sum_j M_{ij}^2 \pi_j
= \left. \frac{\partial^2 V}{\partial \theta_i \partial \theta_j}\right|_0 \pi_j = 
- \sum_j \sum_\alpha
\frac{n_{i,\alpha}\,n_{j,\alpha}}{v_i v_j}\,
\beta_j
\left(\left. V_\alpha\right|_0+\hc\right),
\qquad \forall i \, ,
\end{equation}
which immediately implies the first condition,
\begin{equation}
\sum_j n_{j,\alpha}\,\frac{\beta_j}{v_j}=0 \, .
\end{equation}

The second condition follows from requiring the physical NGB to be orthogonal to the NGBs absorbed by the gauge bosons. The scalar--gauge mixing induced by the covariant derivative is
\begin{equation}
\L_{mix}=-\sum_{a,i} g_a\,A_\mu^a\,t_i^a\,v_i\,\partial_\mu\theta_i \, ,
\end{equation}
where $t_i^a$ denotes the gauge charge associated with the generator $T^a$ acting along the vacuum direction. Orthogonality therefore requires
\begin{equation}
\sum_j t_j^a\,v_j\,\beta_j=0 \, ,
\qquad \forall a \, .
\end{equation}

In summary, the physical NGB can be written as $\pi=\sum_i\beta_i\theta_i$, where the coefficients $\beta_i$ satisfy
\begin{align}
\sum_j n_{j,\alpha}\,\frac{\beta_j}{v_j}&=0 \, , \label{eq:condNGB1}\\
\sum_j t_j^a\,v_j\,\beta_j&=0 \, ,
\qquad \forall a \, .
\label{eq:condNGB2}
\end{align}

These conditions are equivalent to those derived in the classic reference~\cite{Srednicki:1985xd}. In particular, Eq.~\eqref{eq:condNGB1} can be interpreted as the conservation of an exact global $U(1)$ symmetry whose charges define the physical NGB. This becomes manifest by introducing
\begin{equation}
\beta_j \equiv q_j\,v_j \, .
\end{equation}
The two conditions then take the form
\begin{align}
\sum_j n_{j,\alpha}\,q_j&=0 \, , \label{eq:condNGB1q}\\
\sum_j t_j^a\,v_j^2\,q_j&=0 \, ,
\qquad \forall a \, .
\label{eq:condNGB2q}
\end{align}
The first equation is simply the condition of charge conservation for the global symmetry, while the second enforces orthogonality with the gauge NGBs directions.

For the present model, these conditions become
\begin{align}
\sum_j n_{j,\alpha}\,q_j&=0
\quad\Rightarrow\quad
q_z+q_\eta-q_H=0\, ,
\\
\sum_j t_j^a\,v_j^2\,q_j&=0
\quad\Rightarrow\quad
\begin{cases}
q_Hv_H^2+q_\eta v_\eta^2=0\, ,\\
q_Hv_H^2+q_wv_w^2-q_zv_z^2=0\, ,\\
q_\eta v_\eta^2-q_wv_w^2+q_zv_z^2=0\, .
\end{cases}
\end{align}
Solving these equations yields
\begin{equation}
q_w=\frac{v_z^2}{v_w^2}
\left(1-\frac{v_H^2v_\eta^2}{v^2v_z^2}\right)q_z\, ,
\qquad
q_H=\frac{v_\eta^2}{v^2}q_z\, ,
\qquad
q_\eta=-\frac{v_H^2}{v^2}q_z\, ,
\end{equation}
and the corresponding majoron direction is therefore
\begin{align}
J=\frac{1}{v_J}
\begin{pmatrix}
\frac{v_\eta^2}{v^2}v_H &
v_z &
\frac{v_z^2}{v_w^2}
\left(1-\frac{v_H^2v_\eta^2}{v^2v_z^2}\right)v_w &
-\frac{v_H^2}{v^2}v_\eta
\end{pmatrix},
\end{align}
with the normalization fixed by
\begin{equation}
v_J^2=\sum_j q_j^2v_j^2\, .
\end{equation}

Finally, rotating to the basis defined in Eq.~\eqref{eq:scalarbases}, in which the bidoublet components transform irreducibly under $SU(2)_L$, the physical states become
\begin{align}
\P_1
&\simeq
\mathcal{N}_{\P_1}
\left(
-\frac{v_\eta}{v_H}\ \ 
\frac{v_\eta}{\sqrt{2}v_\Phi}\ \ 
-\frac{v_\eta}{\sqrt{2}v_\Phi}\ \ 
1
\right),
&&
M_{\P_1}^2
=
\lambda_\Phi'
\frac{v_\Phi^4}{v_\eta^2}
\delta \, ,
\\
J
&\simeq
\mathcal{N}_J
\begin{pmatrix}
\frac{v_\eta^2}{\sqrt{2}v_Hv_\Phi} &
1 &
\frac{v_\eta^2}{2v_\Phi^2}
\left(
1-\frac{M_\eta^2}{\lambda_\Phi'v_\Phi^2}
\right) &
-\frac{v_\eta}{\sqrt{2}v_\Phi}
\end{pmatrix},
&&
M_J^2=0 \, .
\end{align}
Here $\mathcal{N}_J$ denotes the approximate normalization factor of the majoron state.
\subsection{Charged mass matrix} \label{subsec:charged}

In the basis $\mathcal{S}^{+}$ of Eq.~\eqref{eq:scalarbases}, the charged scalar mass matrix reads
\begin{equation} \label{eq:Mcharged}
\M_{S^+}^2 = \frac{v_z \left(v_w^2 - v_z^2 \right) \lambda_{\Phi}'}{2}
\begin{pmatrix}
-\frac{v_z}{v_H^2} & \frac{1}{\sqrt{2}\,v_H} & \frac{1}{\sqrt{2}\,v_H} & \frac{v_w}{v_H\,v_\eta} \\
\frac{1}{\sqrt{2}\,v_H} & \frac{1}{v_w - v_z} & 0 & \frac{1}{\sqrt{2}\,v_\eta} \\
\frac{1}{\sqrt{2}\,v_H} & 0 & -\frac{1}{2 v_\Phi} & -\frac{1}{\sqrt{2}\,v_\eta} \\
\frac{v_w}{v_H\,v_\eta} & \frac{1}{\sqrt{2}\,v_\eta} & -\frac{1}{\sqrt{2}\,v_\eta} & -\frac{v_z}{v_\eta^2}
\end{pmatrix} \, .
\end{equation}
As expected, it has two vanishing eigenvalues, associated with the NGBs eaten by $W$ and $W'$. The diagonalization of the remaining block proceeds exactly as described in Sec.~\ref{subsec:CPeven}, and leads to the physical states quoted in the main text.
\section{The $\mathcal{S} \, J \, J$ coupling} \label{App:SJJ}
We now derive the trilinear coupling between one CP-even scalar and two majorons. In general, this interaction originates from the scalar potential through the trilinear term
\begin{equation}
    \L_{S A A} = \sum_{ijk} \frac{1}{2} V_{ijk} \, S_i \, A_j \, A_k \, , \qquad V_{ijk} = \left. \frac{\partial^3 V}{\partial S_i \partial A_j \partial A_k}  \right|_0 \, .
\end{equation}
The form of these interactions is constrained by the underlying global $U(1)$ symmetry. In particular, the dependence of the potential on the CP-even and CP-odd fields is related by the symmetry transformation
\begin{align}
    V\left( \Phi_i \,, \, \Phi_i^* \right) = V\left( e^{i q_i \alpha} \Phi_i \,,\, e^{-i q_i \alpha}\Phi_i^* \right) \simeq V\left( \Phi_i + i q_i \alpha \Phi_i \,,\, \Phi_i^* -  i q_i \alpha \Phi_i^* \right). 
\end{align}
Expanding to first order in $\alpha$ gives the identity
\begin{equation}
     \sum_i q_i \left( \Phi_i \, \frac{\partial V}{\partial \Phi_i} -  \Phi_i^* \, \frac{\partial V}{\partial \Phi_i^*} \right) = 0 \, ,
\end{equation}
which can be written in terms of their CP-even and CP-odd components ($\Phi_i \equiv \frac{1}{\sqrt{2}} \left( v_i + S_i + i A_i \right)$),
\begin{equation}
     \sum_i q_i \left[ A_i \, \frac{\partial V}{\partial S_i} -  \left( S_i+v_i \right) \, \frac{\partial V}{\partial A_i} \right] = 0 \, .
\end{equation}
This relation implies general identities among the derivatives of the scalar potential. Differentiating with respect to $A_j$ and $S_k$ yields
\begin{equation}
   q_j \, \frac{\partial^2 V}{ \partial S_k \partial S_j } -  q_k \, \frac{\partial^2 V}{ \partial A_j \partial A_k } -  \sum_i q_i \,S_i  \,  \frac{\partial^3 V}{ \partial S_k \partial A_j \partial A_i } = 0 \, .
\end{equation}
Evaluating this expression in the vacuum we obtain
\begin{equation}
    q_j \, \left(\M_\S^2 \right)_{kj} -  q_k \, \left(\M_\P^2 \right)_{jk}- \sum_i q_i \,v_i  \,  \left. \frac{\partial^3 V}{ \partial S_k \partial A_j \partial A_i } \right|_0 = 0 \, .
\end{equation}
As shown in the Sec.~\ref{subsec:CPodd}, the components of a physical NGB are proportional to $q_i v_i$, where the charges $q_i$ satisfy Eqs.~\eqref{eq:condNGB1q} and~\eqref{eq:condNGB2q} and correspond to a conserved charge. The last term in the previous equation is therefore precisely the contraction of the trilinear tensor $V_{ijk}$ with a NGB direction. Denoting a generic NGB by $\pi$, the trilinear interaction can be written as
\begin{equation}
    \L_{S_i \pi \pi} = \sum_{ijk} \frac{v_\pi^2}{2} V_{ijk} \, q_j \, v_j \,q_k \, v_k \, \S_i \, \pi \, \pi = \sum_{ij}  \frac{v_\pi^2}{2}  \, q_j^2 \, v_j \, \left(\M_\S^2 \right)_{ji}  \, S_i \, \pi \, \pi \, ,
\end{equation}
where we have used the fact that a NGB is an eigenstate of $\M_\P^2$ with zero eigenvalue. Here $v_\pi$ denotes the normalization factor of the NGB state. Rotating to the CP-even mass basis we find,
\begin{equation}
    \L_{\S_n \pi \pi} =\sum_{ij}  \frac{v_\pi^2}{2}  \, q_j^2 \, v_j \, \left(\M_\S^2 \right)_{ki} \, U_{in} \, \S_{in} \, \pi \, \pi =\sum_{j}
    \frac{v_\pi^2}{2}  \, q_j^2 \, v_j U_{jn} \, M_{\S_n}^2  \, \S_n \, \pi \, \pi \, ,
\end{equation}
where $U$ is the matrix that diagonalizes the CP-even scalar sector. The coupling is therefore proportional to the scalar mass squared, as expected from the derivative nature of NGB interactions. 

Applying this result to the decay $h\to JJ$ in the present model, we obtain
\begin{equation}
    \L_{\S_n \pi \pi} = \frac{1}{2}  \frac{\lambda_{H\Phi}\,m_h^2\,v_H}{\left(2\lambda_\Phi+\lambda_\Phi'\right)v_\Phi^2}  \, \S_n \, \pi \, \pi \, .
\end{equation}

\bibliographystyle{utphys}
\bibliography{mybib.bib}

\end{document}